# Modulations of the photoswitching performances of azobenzene-based molecular devices by the π-conjugated spacers.

Laurette-Apolline Jerro,[1,§] Imen Hnid,[1,§] Imane Arbouch,[2] Ali Yassin,[3,4] David Guérin,[1]
Alexis Elias,[2] Colin van Dyck,[5] Lionel Sanguinet,[3] Jérôme Cornil,[2] Stéphane Lenfant,[1]
Philippe Blanchard[3] and Dominique Vuillaume.[1*]

*1) Institute for Electronics Microelectronics and Nanotechnology (IEMN), CNRS, University of Lille, Av. Poincaré, Villeneuve d'Ascq, France.*
*2) Laboratory for Chemistry of Novel Materials, University of Mons, Belgium.*
*3) MOLTECH-Anjou, CNRS, SFR MATRIX, University of Angers, Angers, France.*
*4) Department of Physical Sciences, School of Arts and Sciences, Lebanese American University, Beirut, Lebanon.*
*5) Theoretical Chemical Physics group, University of Mons, Mons, Belgium.*

** Corresponding authors: dominique.vuillaume@iemn.fr*
§ *These authors* contributed equally.

## Abstract.

The nature of the π-conjugated spacer is a key element to tune the electrical conductance and optical switching properties of metal-spacer-azobenzene-metal molecular devices. We report that the azobenzene-based molecular junctions with a thiophene or a phenylthiophene spacer exhibit a high *cis/trans* conductance ratio (ca. $10^3$), while those with a bithiophene spacer have a lower conductance switching ratio (ca. 500) at the stated conditions (light wavelength and intensity) and for one batch of samples (out of two). Similarly, the two former molecular junctions have quite similar *trans*-to-*cis* isomerization kinetics, while the latter has a lower switching rate. The energy level of the highest occupied molecular orbital, which controls the electron transfer in the molecular junction, is weakly influenced by the nature of the spacer, nor by the conformation of the azobenzene moieties. The large increase in the conductance of the molecular junction with azobenzene in the *cis*-state is explained by an increase in the electronic coupling energy between the molecule and the top electrode. Theoretical studies by Density Functional Theory and Non Equilibrium Green's Functions are qualitatively consistent with the experiments. We find a consistent structure-property trend, showing that the higher molecule/electrode coupling energy for the *cis*-state is explained by the top benzene ring of the azobenzene lying flat with respect to the top electrode for the thiophene, phenylthiophene and bithiophene spacers. On the contrary, we observe that a spacer with a carbon-carbon triple bond (acetylene thiophene) induces a large irreproducibility and unpredictability of the results.

## 1. INTRODUCTION.

Azobenzene is an archetype of photochromic molecules that was widely studied to design materials and devices with light-driven mechanical, thermal, optical, electron transport, and bioadaptive properties.[1-5] At the molecular scale, when connected between two electrodes in a molecular junction (MJ), these devices show a light-triggered switch between a low electronic conductance and a high electronic conductance state.[1, 6-8] Whether in a single molecule or self-assembled monolayers (SAM)-based molecular junction, the azobenzene moiety is connected to the electrodes by a spacer. In most of the cases, this spacer is a short alkyl chain,[9-13] a π-conjugated monomer (phenyl most of the time)[14, 15] or a combination of the two (e.g., butyl-bithiophene).[16] This spacer is used to decouple the molecule from the metallic substrate to avoid the quenching of the excited state and to better accommodate the conformation change upon photoisomerization.[17] However, the drawback of the alkyl spacers is to reduce the overall conductance of the MJs. Conversely, the effect of a π-conjugated spacer (beyond a simple phenyl group) on the electronic properties of the MJs and their photo-switching performances have not been studied in depth. In a recent work,[18] the *cis*/*trans* conductance ratios of a series of oligophenylene-azobenzene derivatives (with 1, 2 or 3 phenyl rings between the thiol anchor and the azobenzene moiety) were calculated by density functional theory (DFT) and no significant correlation was predicted between the *cis*/*trans* conductance ratios and the number of phenyl groups (the STM break junction experiments failed to measure these ratios because the single molecule conductances in the *trans*-state were too low). The electronic coupling strength between the azobenzene moiety and the electrode, through the spacer, is expected to play a key role in controlling the electron conductance and the conductance switching properties.

In this work, we decipher how the nature of the π-conjugated spacer between a gold electrode and an azobenzene moiety tunes the electronic and switching properties of the metal-spacer-azobenzene-metal MJ. We synthesized 4 molecules (scheme 1), one primary system with a thiophene spacer (molecule AZ1), followed by three other molecules with a more extended spacer, *i.e*. bithiophene (AZ2), phenylthiophene (AZ3), and acetylene thiophene (AZ4). The electron transport (ET) properties of the MJs were studied by forming a self-assembled monolayer (SAM) on an ultra-flat, template-stripped, gold electrode ($^{TS}$Au) gently contacted by the tip of a conductive-AFM (C-AFM), in the following referred to as $^{TS}$Au-S-AZn/tip (where "-" denotes a chemical bond, "/" a mechanical contact and AZn with n=1 to 4 indicating the corresponding azobenzene derivative). Hundreds of current-voltage (I-V) curves were acquired for each of the 4 MJs to determine how the spacer tunes the electron conductance ratio between the two isomers. These data sets were statistically analyzed by a single-energy level (SEL) analytical model[19, 20] based on the Landauer-Büttiker-Imry formalism[21, 22] and the molecular orbital dominating the ET was determined for both the *trans* and *cis* isomers of the 4 MJs. We performed I-V measurements at different times of light irradiation to determine the photoisomerization switching times of the MJs (*trans* to *cis* and *cis* to *trans*).

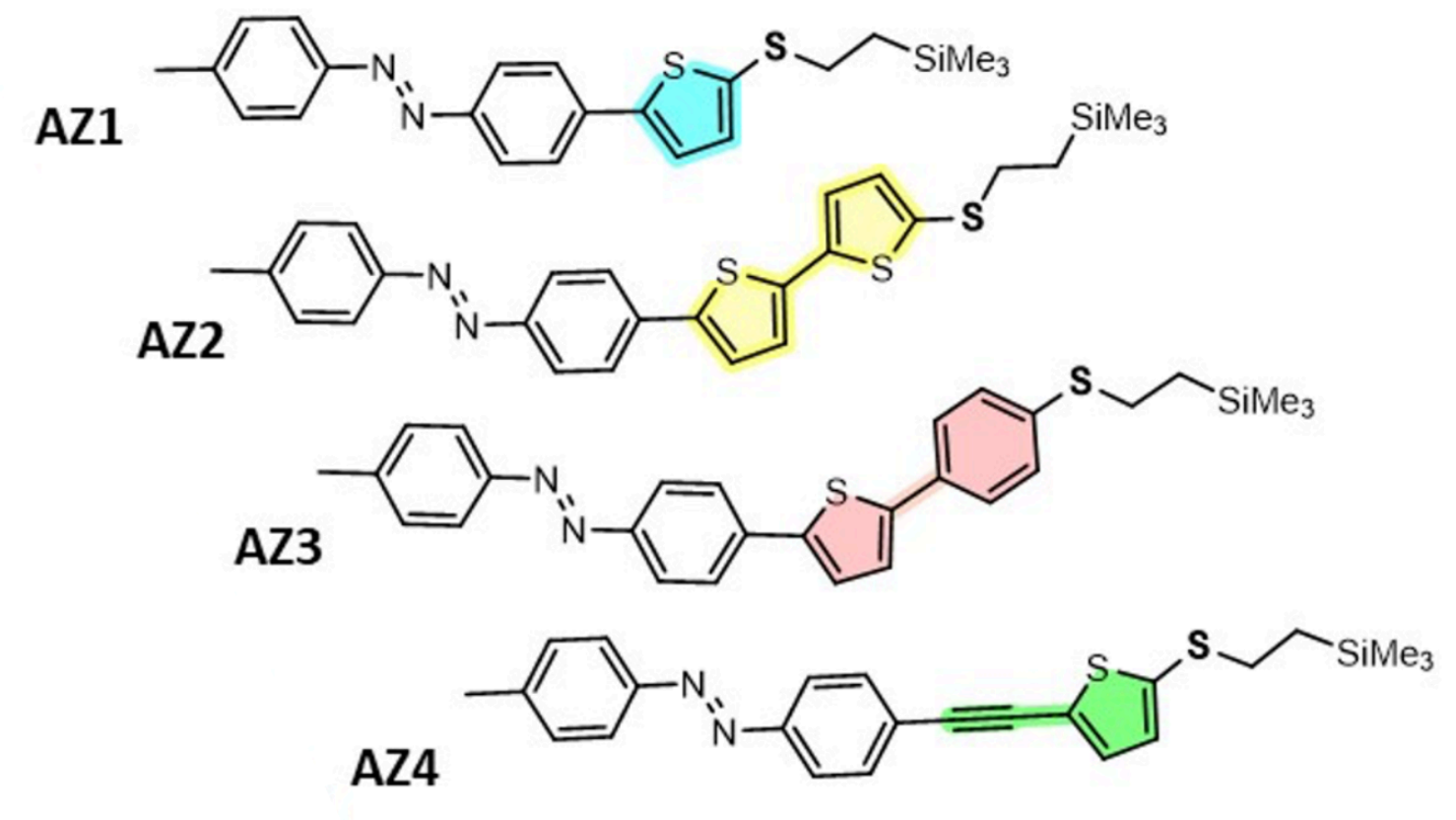

***Scheme 1**. Chemical structures of the 4 azobenzene molecules with different spacers (highlighted in color). The molecules are equipped with a sulfur-protected (TMSE, trimethylsilyl-ethyl group) for the formation of self-assembled monolayers on Au substrates.*

All the 4 MJs show higher electronic conductances in the *cis* state than in the *trans*-one. We found that the AZ1 and AZ3 MJs present the highest *cis/trans* conductance ratios (≈ $10^3$), while the AZ2 MJ exhibits a lower ratio (≈ 500) and the back conversion for *cis* to *trans* isomers can only occur thermally. The behavior of the AZ4 MJ is specific. Introducing a spacer with a carbon-carbon triple bond (acetylene thiophene) leads to experimentally irreproducible (batch-to-batch inconsistency) and unpredictable performances by theory. The ET is dominated by the energy level of the HOMO, which is weakly dependent on the nature of the π-spacer, nor strongly modified upon *trans*/*cis* isomerization. The huge increase in current for the MJs with the *cis* isomer is ascribed to a better electronic energy coupling between the molecules and the electrodes. The study was completed by Density Functional Theory (DFT) coupled with the Non Equilibrium Green's Function (NEGF) formalism to describe the electronic properties of the junction at zero bias. We report a qualitative agreement for the trends found for the cis/*trans* conduction ratios for the AZ1-AZ3 MJs, the variations of the HOMO energy levels and the electronic molecule/electrode coupling energies. The AZ1, AZ3 and AZ4 MJs also have quite similar *trans*-to-*cis* switching kinetics, while the AZ2 MJ is less efficient (lower switching rate). The AZ3 MJ exhibits the slowest light-driven back-switching from *cis*-to-*trans* isomers.

## 2. EXPERIMENTAL RESULTS.

### *2.1. Molecules and self-assembled monolayers.*

Azobenzene derivatives AZ1–AZ3 were synthesized via a specific Pd-catalyzed C–H direct arylation method starting from azobenzene derivative **1** (scheme 2). Separately to this brominated derivative, trimethyl(2-(thiophen-2-ylthio)ethyl)silane (**2**) was prepared from commercially available thiophene-2-thiol following reported synthesis.[23] Similarly, the trimethylsilyl (TMS)-protected building blocks (**3**) and (**4**), based on 2,2'-bithiophene and 2-phenylthiophene respectively, were likewise obtained according to literature.[24] In our optimized procedure, typically oven-dried microwave tubes are charged with derivative **1** (1 eq.), followed by either of the corresponding spacers, **2**,

**3**, or **4**, (1 eq.) palladium acetate (0.1 eq.), tricyclohexylphosphine tetrafluoroborate (0.2 eq.), pivalic acid (1 eq.) and $K_2CO_3$ (3 eq.) in dry toluene under argon atmosphere. The reaction mixtures were exposed to microwave irradiation for 2 hours at 150°C. The obtained residues were all purified by column chromatography on silica gel affording the corresponding products as orange solids. AZ4, on the other hand, was synthesized using a different synthetic approach, also starting from azobenzene derivative **1**. The latter was engaged in a Stille type reaction with palladium catalyst to introduce an acetylenic bond in para position relative to the azo group. Subsequent deprotection of the trimethylsilyl (TMS) group using tetra-n-butyl ammonium fluoride ($Bu_4NF$) as a source of fluoride ion generated key derivative **5**. In parallel, a fresh batch of iodinated thiophene derivative **6** was prepared and then directly engaged with **5** in a Sonogashira type reaction overnight to yield AZ4 as a shiny orange solid. Details on the synthetic procedures and physicochemical characterizations (NMR, HRMS) for compounds AZ1–AZ4 are given in the Supporting Information. Figure S2 and Table S1 present the UV-Vis absorption spectra of the four molecules recorded in DCM solutions, and report the wavelengths at the maximum of their optical absorption $\lambda_{max}$. These values were used to define the wavelengths used to induce the photo-isomerization of the MJs. It is noted that the four systems display their major absorption bands in the 300–500 nm region.

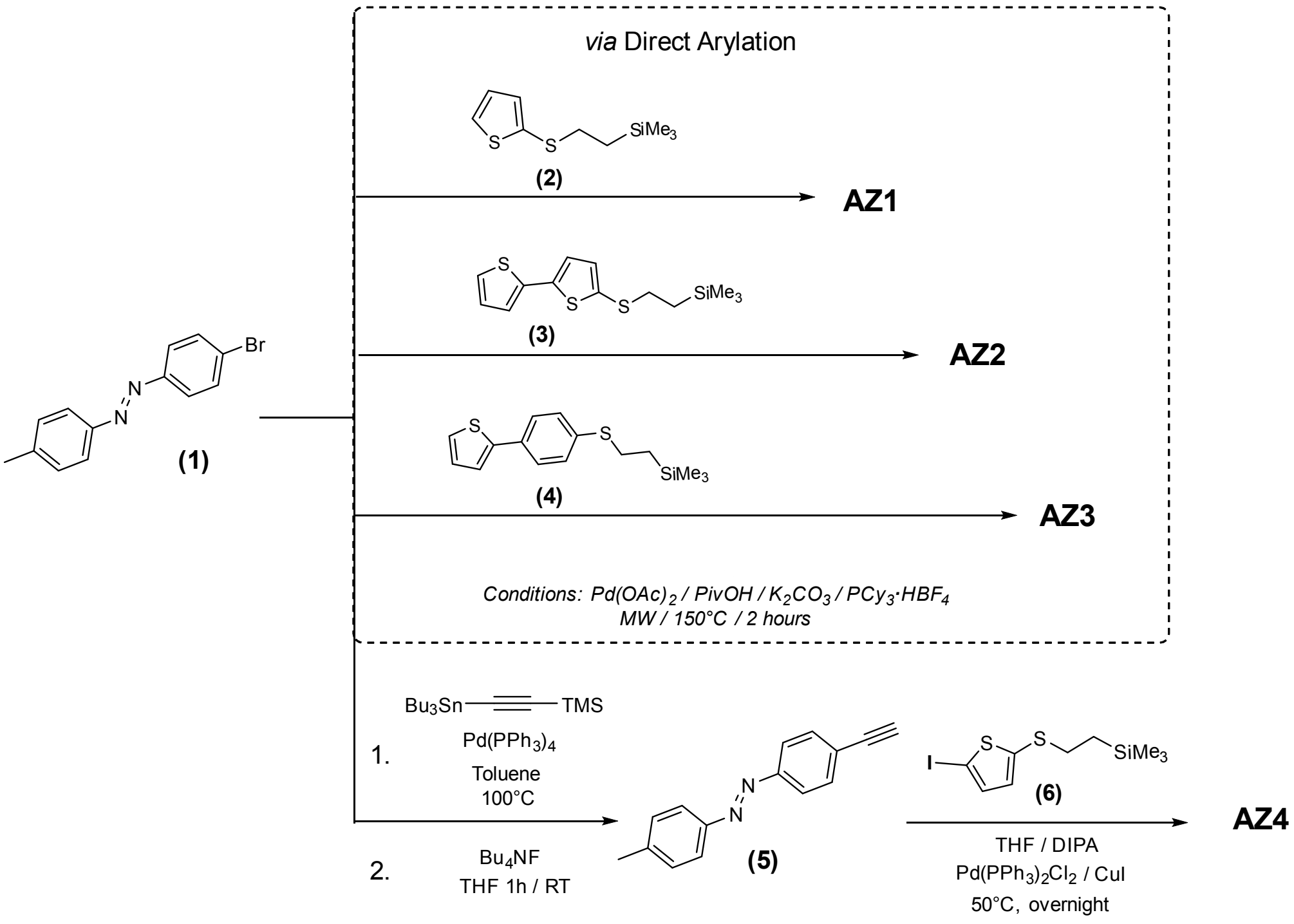

***Scheme 2.*** *Synthetic pathways to AZ1, AZ2, and AZ3 via direct arylation, and AZ4 via successive Stille and Sonogashira reactions.*

The $^{TS}$Au-S-AZn/tip MJs are constituted by forming SAMs on $^{TS}$Au substrate following standard protocols (see details in the Supporting Information). Two batches were fabricated with the same protocol. The quality of the SAMs was assessed by measuring the SAM thickness (spectroscopic ellipsometry). The measured thicknesses are smaller than the theoretical length of the molecules (in their *trans* isomer), see Table S2 in the Supporting Information, and we deduced tilt angles of the long axis of the molecules with the surface normal of ≈ 43-61° (Table S2).

***2.2. Conductance switching: Conductive-AFM measurements.***

The SAMs were contacted by the PtIr tip of a C-AFM to acquire hundreds of I-V curves with the azobenzene in the *trans* and *cis* isomers. In the following, we systematically present the results for batch #2. In several cases we compared with data for batch #1 when appropriate; all data for batch #1 are given in the supporting information. Figure 1 shows the I-V data sets of the four samples in the pristine state just after the SAM formation (we assume that the azobenzenes are mainly, but not totally, in the *trans* isomer, *vide infra*), after light irradiation at 385-405 nm to switch the azobenzenes in their *cis* isomer and back to the *trans* isomer after light irradiation at 325-365 nm or thermal back conversion in the case of AZ2 (see details in panels of Fig. 1). Figure 2 shows the statistical distribution of the currents at + and - 0.75 V.

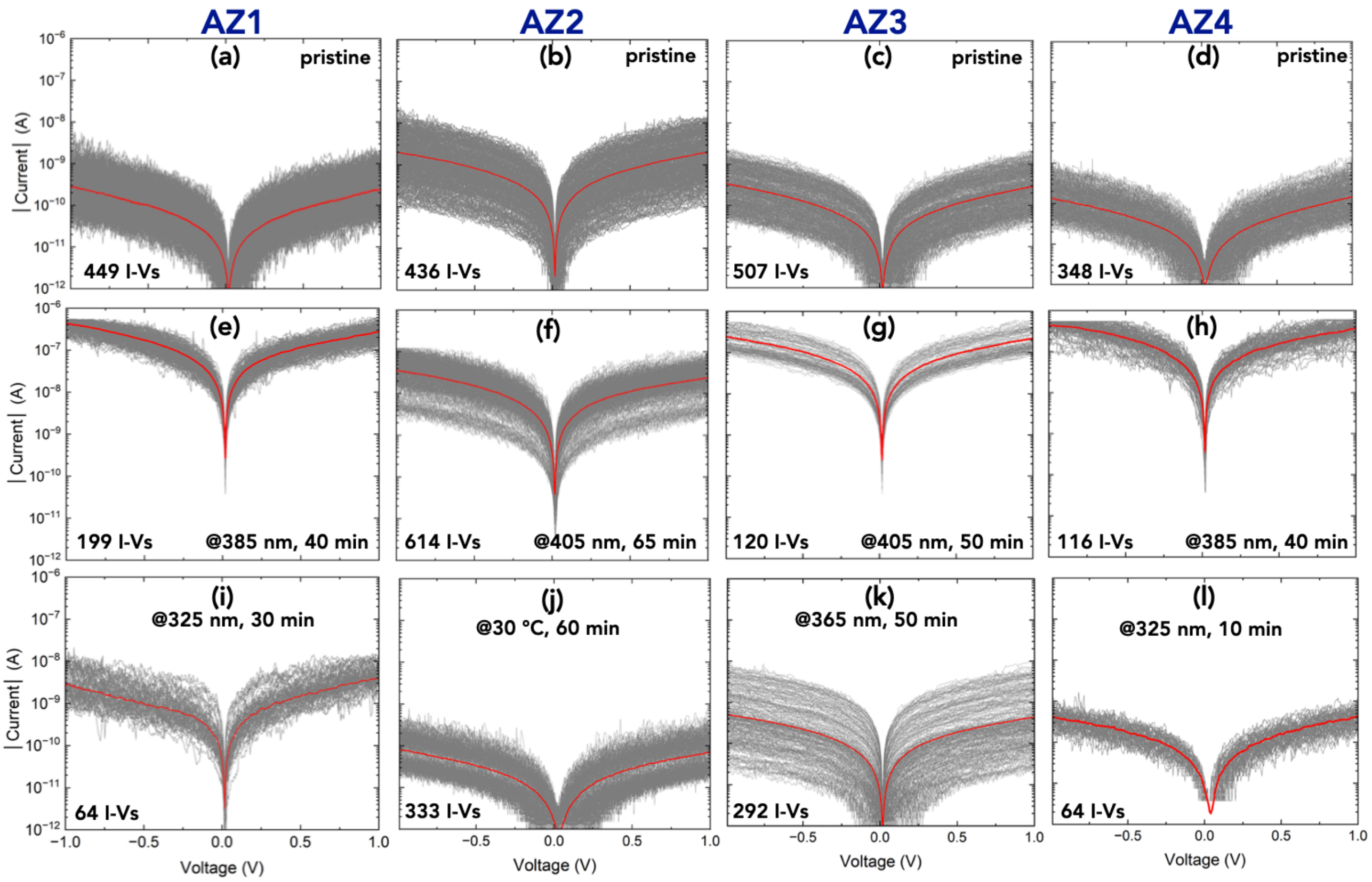

***Figure 1.*** *Current-voltage (I-V) data sets for the four* $^{TS}$*Au-S-AZn/tip MJs (n=1 to 4, batch #2) :* ***(a-d)*** *pristine MJs after fabrication,* ***(e-h)*** *after light irradiation (385-405 nm) to induce trans-to-cis isomerization and* ***(i-l)*** *after light irradiation (325-365 nm) to induce the cis-to-trans isomerization (except for AZ2, thermal back isomerization). The red lines are the mean current-voltage ($\bar{I}$-V) curves.*

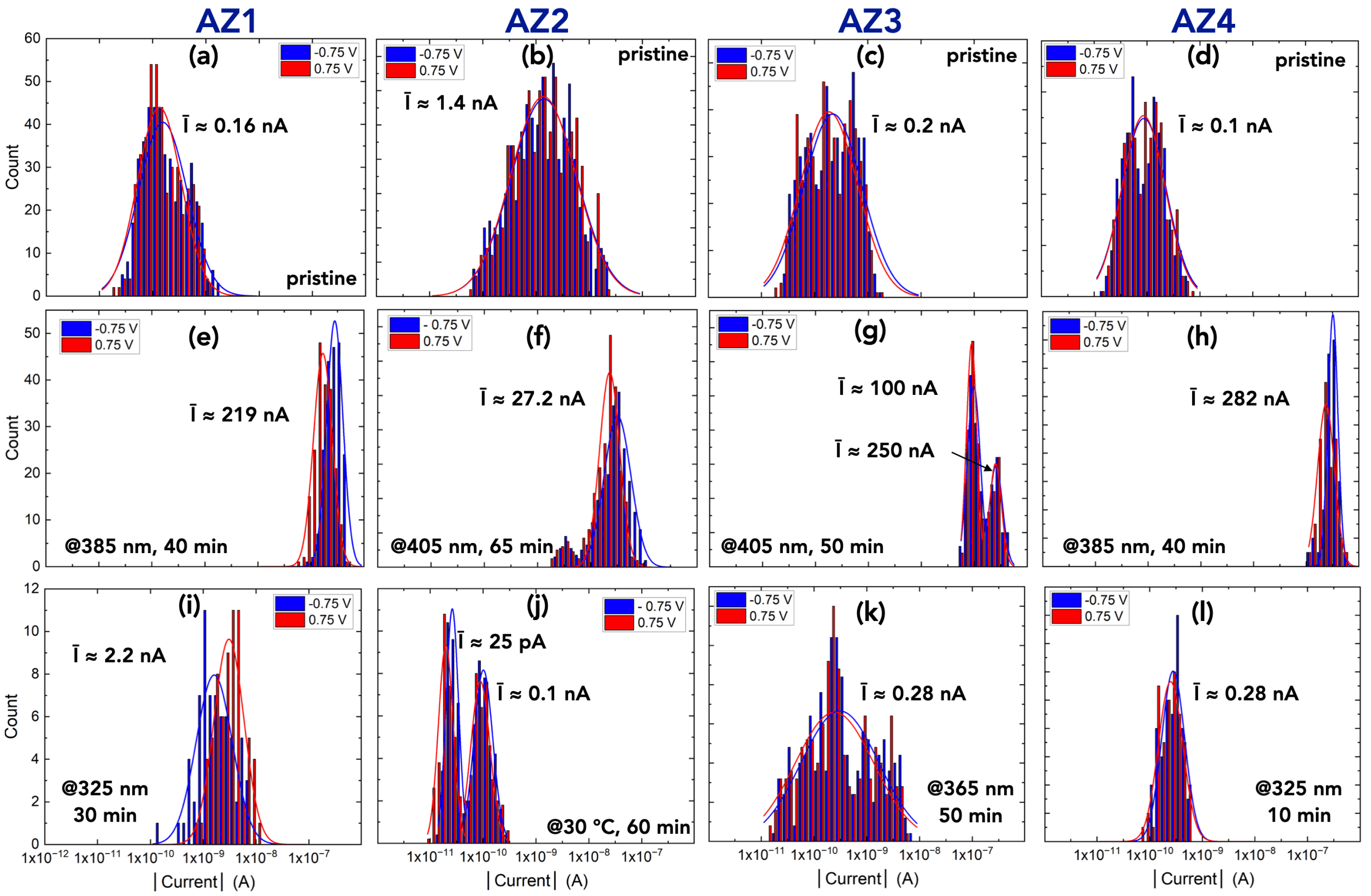

***Figure 2.*** *Histograms of the current at + and - 0.75V taken from data in Fig. 1 (batch #2). The data are fitted with log-normal distributions (solid lines). The fit parameters of the log-normal distributions of the measured currents given as log-µ ± log-σ (log µ is the mean value of the Gaussian distribution in the decimal log scale, log-σ is the standard deviation in the log scale) are summarized in Table 1. The corresponding mean currents Ī are indicated in the panels. When a small voltage asymmetry is observed, the values reported in the panels are the average values between the values at positive and negative voltages. In two cases, a double distribution is clearly observed.* ***(a-d)*** *pristine MJs after fabrication,* ***(e-h)*** *after light irradiation (385-405 nm) to induce trans-to-cis isomerization and* ***(i-l)*** *after light irradiation (325-365 nm) to induce cis-to-trans isomerization (except for AZ2, thermal back isomerization).*

A reversible *trans/cis* switching of the azobenzene moiety is observed for the four $^{TS}$Au-S-AZn/tip MJs. Except for the $^{TS}$Au-S-AZ2/tip MJ, the mean currents in the pristine state after SAM formation are in the same range (≈ 0.1-0.2 nA at ± 0.75 V). For the AZ2, the mean current is higher (≈ 1.4 nA at ± 0.75 V), with a slightly larger broadening of the distribution. After light irradiation (385-405 nm) to induce *trans*-to-*cis* isomerization, AZ1, AZ3 and AZ4 MJs display the highest current (≈ 0.1-1 µA at ± 0.75 V). After light irradiation (325-365 nm) to induce the back *cis*-to-*trans* isomerization (except for AZ2, thermal back isomerization), the currents are reduced but with behaviors that are sample dependent. The AZ4 MJ has recovered almost the same current level as in the pristine state (Fig. 2). It is also almost the case for the AZ3 MJ, but with a larger dispersion of the data set. In the case of the AZ1 MJ, the recovery is not complete (but a shorter irradiation time was used, *vide infra* the study of switching kinetics). Again, the case of AZ2 MJ is specific, with a current much smaller than the pristine state after the thermal treatment (likely due to a mix of *trans* and *cis* isomers in the pristine SAM) and a clearly distinguishable double distribution. Table 1 summarizes the values of the mean current $\bar{I}$ measured at ± 0.75 V (from Fig. 2) and the SAM zero-bias conductance G(0) (from the first derivative ∂I/∂V of the mean $\bar{I}$-V - Fig. 1 - taken in the voltage window ± 100 mV). We calculated the ratios between the *cis* and pristine states $R_{cis/0}$ after light irradiation (385-405 nm) to induce the *trans*-to-*cis* isomerization and the *cis/trans* conductance ratios $R_{cis/trans}$ after a second light irradiation (325-365 nm) to induce the *cis*-to-*trans* isomerization (except for AZ2, thermal back isomerization). The ratios $R(I)_{cis/0}$and $R(I)_{cis/trans}$ were calculated from the mean current $\bar{I}$ at ± 0.75 V determined from the statistical analysis shown in Fig. 2. Since these ratios are voltage dependent (Fig. S6), we also calculate ratios $R(G)_{cis/0}$ and $R(G)_{cis/trans}$ from the SAM zero-bias conductance G(0). Note that $R(I)_{cis/trans}$ and $R(G)_{cis/trans}$ for AZ1 are underestimated (not complete recovery) as well as $R(I)_{cis/0}$ and $R(G)_{cis/0}$ for AZ2 (slightly higher current in the pristine SAM, likely due to a mix of *trans* and *cis* molecules, *vide supra*). The highest conductance ratios (≈ 2820) are observed for the MJs with the acetylene thiophene spacer (molecule AZ4), followed by the MJs with the thiophene (AZ1, ≈ 1370, albeit a complete back isomerization was not observed), and then the phenylthiophene spacer (AZ3, ≈ 875), Table 1. With the bithiophene spacer (AZ2), we obtained the smallest ratio of the 4 MJs (≈ 450). We observed the same trends (Fig. 3, Table 2) from the zero-bias conductance ratios $R(G)_{cis/0}$ and $R(G)_{cis/trans}$. However, since the light irradiation conditions were not strictly similar (not the same light irradiation time), we more precisely calculated these ratios at the same number of photons from the isomerization kinetic studies (*i.e.*, at $7.6 \times 10^{18}$ photons, Fig. 4-b, *vide infra*). The same trend is observed, with AZ1 and AZ4 showing the best switching ratios (Table 2, Fig. 3-b). The value of ≈ 1 for AZ2 is due to the fact that t*rans*-to-*cis* isomerization has not yet significantly started with this number of photons (*vide infra*, the kinetic study). Also note that the *cis* to *trans* back isomerization for AZ2 was only possible thermally, as also observed in solution. For the samples of batch #1, we have observed qualitatively a similar reversible switching behavior, albeit with some variability of the measured current levels and switching ratios (see data sets in the Supporting Information, Figs. S7 and S8, summarized in Tables 1 and 2). Fig. 3-a compares the variations of the zero-bias conductance of the 4 samples in the *trans* and *cis* states. The behaviors are quite similar for the two batches, except for the AZ4 MJs that do not display the highest ON/OFF ratio as for the AZ4 MJs in batch #2. We also note that the levels of conductances in the *cis*-state are lower for batch #1, while almost similar in the *trans*-state, leading to smaller switching ratios, as summarizes in Fig. 3-b. This feature likely reflects the batch-to-batch variability of the SAMs.

| | batch #2 | | | | | batch #1 | | | | |
|---|---|---|---|---|---|---|---|---|---|---|
| | log-μ ± log-σ<br>$I_0$ (nA) | log-μ ± log-σ<br>$I_{cis}$ (nA) | log-μ ± log-σ<br>$I_{trans}$ (nA) | $R(I)_{cis/0}$ | $R(I)_{cis/trans}$ | log-μ ± log-σ<br>$I_0$ (nA) | log-μ ± log-σ<br>$I_{cis}$ (nA) | log-μ ± log-σ<br>$I_{trans}$ (nA) | $R(I)_{cis/0}$ | $R(I)_{cis/trans}$ |
| AZ 1 | -9.80±0.43<br>0.16 | -6.66±0.17<br>219 | -8.66±0.31<br>2.2 | 1370 | 99.5* | -9.00±0.89<br>1 | -7.51±0.36<br>31 | -8.70±0.65<br>2 | 31 | 15.5 |
| AZ 2 | -8.85±0.61<br>1.4 | -7.56±0.22<br>27.2 | -10.60±0.13<br>-10.00±0.18<br>0.06 | 19.4** | 453 | -10.01±0.68<br>0.1 | -8.79±0.80<br>1.6 | -10.3±0.55<br>0.05 | 16 | 32 |
| AZ 3 | -9.70±0.54<br>0.2 | -7.00±0.18<br>-6.60±0.12<br>175 | -9.55±0.74<br>0.28 | 875 | 625 | -9.82±0.65<br>0.15 | -7.70±0.43<br>20 | -10.36±0.33<br>-9.3±0.34<br>0.27 | 133 | 74 |
| AZ 4 | -10.00±0.41<br>0.1 | -6.55±0.13<br>282 | -9.55±0.22<br>0.28 | 2820 | 1007 | -8.99±0.46<br>1 | -7.8±0.31<br>16 | -8.74±0.39<br>1.8 | 16 | 8.9 |

***Table 1**. Values of the mean current (average of the values at + and - 0.75 V, from the statistical distributions shown in Fig. 2 batch #2 and Fig. S8 batch #1) in the pristine state ($I_0$), after light irradiation to induce trans-to-cis isomerization ($I_{cis}$) and after light irradiation (except for AZ2, thermal back isomerization) to induce cis-to-trans isomerization ($I_{trans}$). The table summarizes the parameters of the log-normal distributions of the measured currents (Figure 2 for batch #2 and Fig. S8 for batch #1) given as log-μ ± log-σ (log μ is the mean value of the Gaussian distribution in the decimal log scale, log-σ is the standard deviation in the log scale). Values in the second line in every row are the corresponding mean values of I in nA given by $10^{log\text{-}\mu +9}$. The current switching (or ON/OFF) ratios between the three states are obtained as $R(I)_{cis/0}=I_{cis}/I_0$ and $R(I)_{cis/trans}=I_{cis}/I_{trans}$ from the mean currents at + and - 0.75 V. Note that when a double distribution is observed (Figs. 2-g, 2-j, S8-k), the value reported in this table is the average of the values of the two peaks. * Value underestimated due to the incomplete cis-to-trans back isomerization (shorter illumination time). ** Value underestimated due to the relatively high currents of the AZ2 pristine SAMs (likely due to a mix of trans and cis AZ2, see text).*

| | batch #2 | | | | | | batch #1 | | | | |
|---|---|---|---|---|---|---|---|---|---|---|---|
| | $G_0$ (nS) | $G_{cis}$ (nS) | $G_{trans}$ (nS) | $R(G)_{cis/0}$ | $R(G)_{cis/trans}$ | $R(G)_{cis/0}$<br>(at $7.6 \times 10^{18}$ ph) | $G_0$ (nS) | $G_{cis}$ (nS) | $G_{trans}$ (nS) | $R(G)_{cis/0}$ | $R(G)_{cis/trans}$ |
| AZ1 | 0.15 | 251 | 2.5 | 1673 | 100* | 1673 | 0.9 | 32.8 | 1.6 | 36 | 20.5 |
| AZ2 | 1.5 | 24.3 | 0.05 | 16.2** | 486 | 1.1*** | 0.16 | 1.3 | 0.15 | 8.1 | 8.7 |
| AZ3 | 0.19 | 153 | 0.36 | 805 | 425 | 360 | 0.18 | 29.8 | 0.25 | 165 | 119 |
| AZ4 | 0.11 | 260 | 0.26 | 2364 | 1000 | 2364 | 0.85 | 13.8 | 1.6 | 16.3 | 8.6 |

***Table 2**. Mean values of the zero-bias conductance (from the first derivative ∂I/∂V of the mean $\bar{I}$-V - Fig. 1 batch #2 and Fig. S7 batch #1 - taken in the voltage window ± 100 mV) in the pristine state ($G_0$), after light irradiation to induce*

*trans-to-cis isomerization ($G_{cis}$) and after light irradiation (except for AZ2, thermal back isomerization) to induce cis-to-trans isomerization ($G_{trans}$). The conductance switching (or ON/OFF) ratios between the three states are obtained as $R(G)_{cis/0}=G_{cis}/G_0$ and $R(G)_{cis/trans}=G_{cis}/G_{trans}$. For batch #2, the last column gives the ratios calculated from the zero-bias conductance measured from the I-V for samples irradiated with the same number of photons. * Value underestimated due to the incomplete cis-to-trans back isomerization (shorter illumination time). ** Value underestimated due to the relatively high currents of the AZ2 pristine SAMs (likely due to a mix of trans and cis AZ2, see text).*** At this number of photons, the isomerization has not yet significantly started (see Fig. 4).*

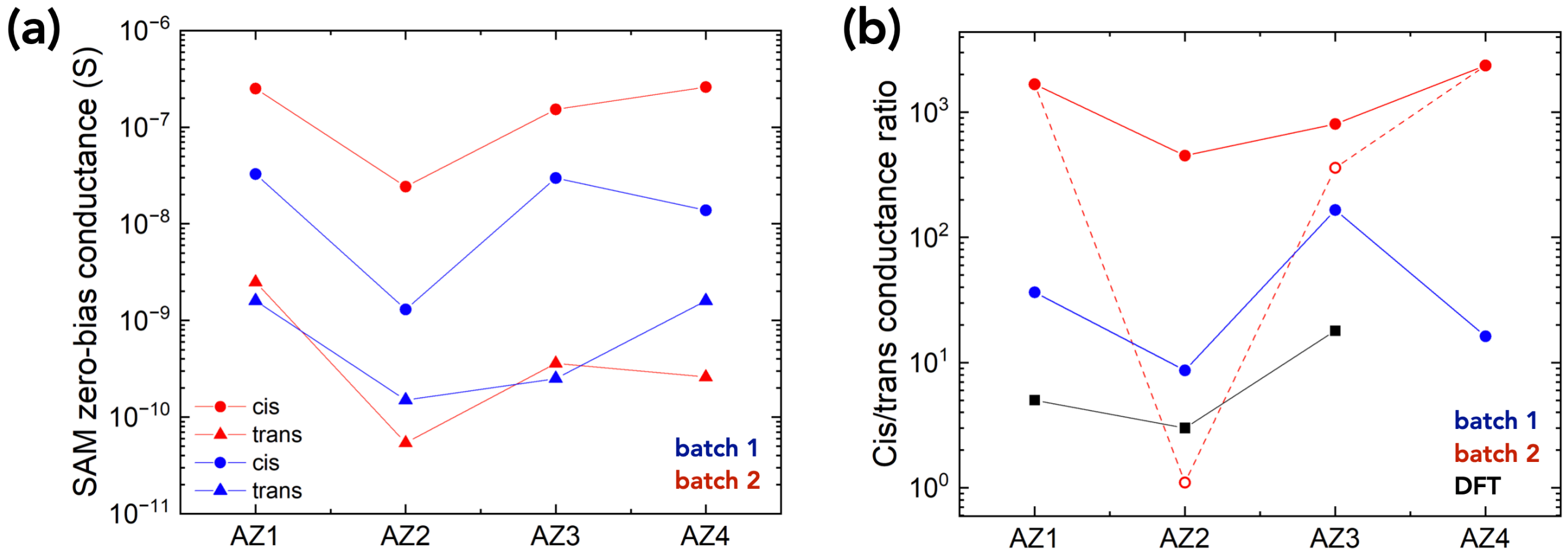

***Figure 3. (a)*** *Zero-bias conductance G(0) of the SAMs with the azobenzene in the cis and trans states for the 4 samples of the batches (data in Table 2).* ***(b)*** *Ratio of the zero-bias conductances in the cis and trans states for the samples of the two batches. In order to compare the best performances of the two batches (while not obtained strictly in the same conditions), the solid circles are the max of $R(G)_{cis/0}$ and $R(G)_{cis/trans}$ (from table 2) for the two batches. Open red circles are the conductance ratios at a constant number of photons ($7.6\times10^{18}$) - only for batch #2. Dark squares are the calculated (DFT) ratio of the electron transmission probability at the Fermi energy (zero bias), Table 3.*

### ***2.3. Isomerization kinetics.***

We studied the switching kinetics (only samples from batch #2) by measuring the I-V curves during step-by-step light irradiation (one step = 10 min). Notice that the measured current is an indirect estimation of the isomer population in the SAMs, and that a more detailed study of the isomerization kinetics would have required a more direct method, *e.g.*, optical absorption that gives a signature of the *trans* or *cis* isomers (shift of the absorption peak). Figure 4-a shows the evolution of the mean current $\bar{I}$ at ± 1V versus the light irradiation time (see the $\bar{I}$-V data sets in the Supporting Information, Fig. S11). From the light intensity measured for the 4 light sources (see the Supporting Information), we replotted the data in Fig. 4-b versus the number of photons, for the purpose of comparison. We observed that both the *trans*-to-*cis* and *cis*-to-*trans* isomerizations roughly follow power law kinetics, indicating a broad dispersion of the time scale of the molecule switching in the SAMs. From the power law exponents (slope of the plots, values in the caption of Fig. 4), we conclude that the AZ1 and AZ4 MJs have the most efficient *trans*-to-*cis* isomerization, followed by AZ3, while AZ2 is the slowest system (Fig. 4-b). For *cis*-to-*trans* isomerization, AZ4 MJs show the fastest switch (recovery of the initial state after ~2-3x$10^{17}$ photons, Fig. 4-c), followed by AZ1 and AZ3 MJs.

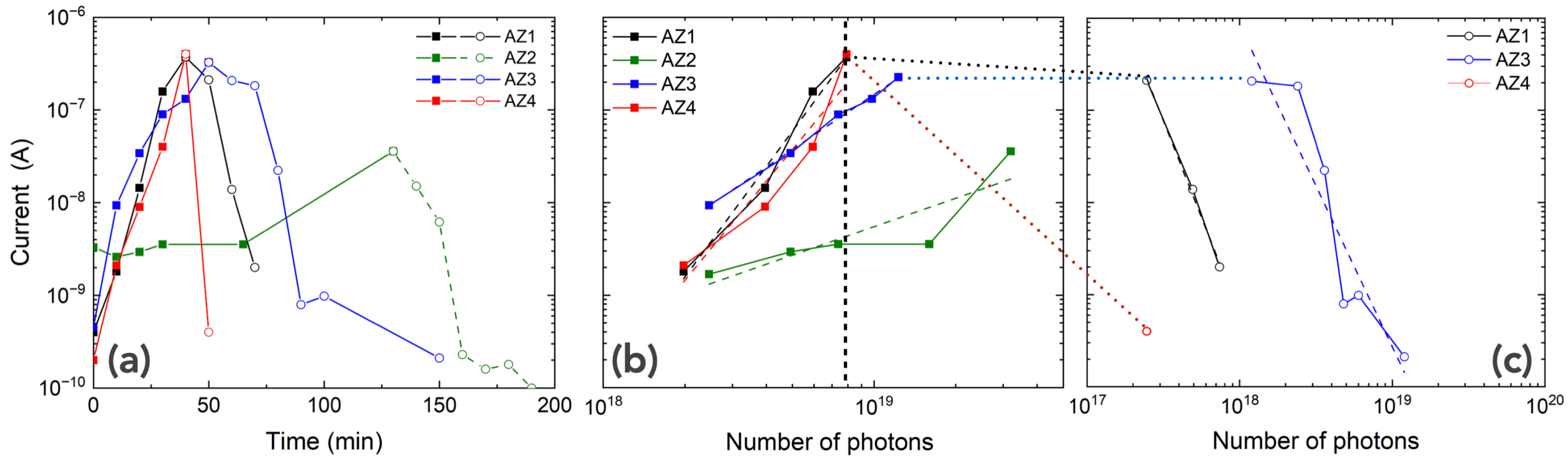

***Figure 4. (a)*** *Evolution of the mean current (average of the values at + and - 1 V) versus the light irradiation time (except AZ2 for the cis-to-trans isomerization: time heating at 30°C - dashed green line). Full (open) symbols for the trans-to-cis (cis-to-trans) isomerization kinetics, respectively.* ***(b)*** *Same data versus the number of photons for trans-to-cis isomerization. The number of photons is the photon flux (measured by a calibrated light power meter, see Supporting Information) multiplied by the light irradiation time (in seconds).* ***(c)*** *Same data versus the number of photons for cis-to-trans isomerization. The dashed lines are the fits with power laws. The dotted lines are a guide for eyes to connect the last data of the panel (b) and the first data of the panel (c). The empiric power law exponents for trans-to-cis isomerization (slope of the log-log plot) are 3.97 ± 0.41, 1.02 ± 0.34, 1.97 ± 0.07 and 3.56 ± 0.79 for AZ1, AZ2, AZ3 and AZ4, respectively (see Table S4 in the Supporting Information). They are -4.21 ± 0.22 and -3.49 ± 0.71 for the cis-to-trans isomerization for AZ1 and AZ3, respectively (Table S4). All data for samples of batch #2.*

### *2.4. Energetics of the molecular junctions.*

We fit the I-V data sets with an analytical single energy level (SEL) model to determine the energy position of the molecular orbital mediating the electron transfer between the electrodes through the molecules (see the Supporting Information). Here, the ET is mediated by the HOMO that is the molecular orbital closest to the Fermi energy of the electrodes $\varepsilon_F$ (see theory section). By fitting the SEL model on every I-V trace of the data sets, we determine the statistical distribution of the energy position of the HOMO, $\varepsilon_0$, with respect to $\varepsilon_F$ for the four $^{TS}$Au-S-AZn/tip MJs with the azobenzene molecules in the *trans* and *cis* isomers. Figure 5 shows the energy level histograms obtained from all the I-V data sets shown in Fig. 1. The general trend is that there is a main peak, which is only slightly affected by changes in conformation (the variations upon *trans*/*cis* isomerization are smaller than or in the same range as the dispersion of the experimental values). In the pristine state, $\varepsilon_0$ is in the range of 0.77-0.81 eV, slightly increasing to 0.85-1.1 eV upon *trans-to- cis* isomerization (Figs. 5 and 7).

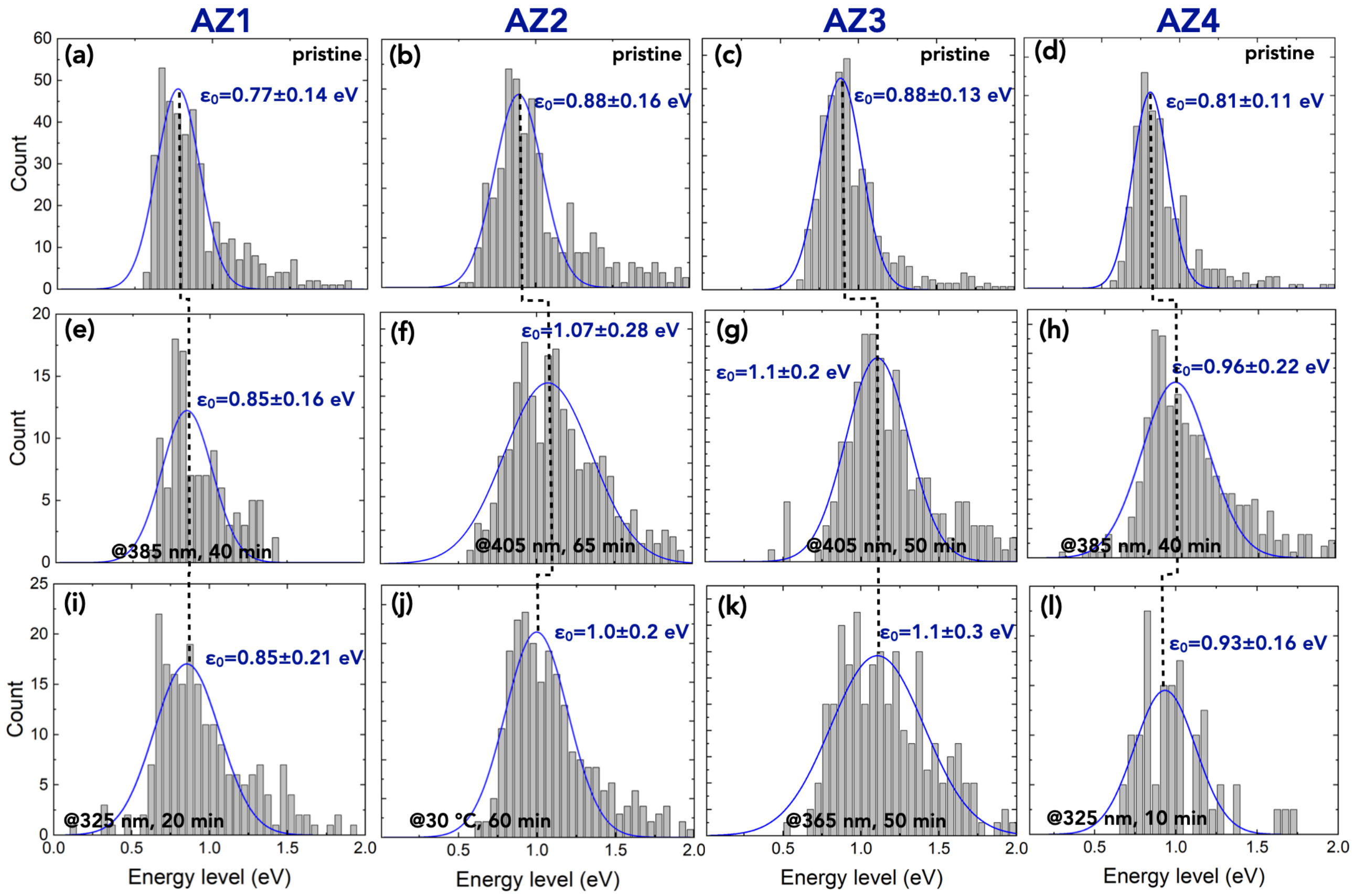

***Figure 5***. *Histograms of the energy level involved in the ET though the four $^{TS}$Au-S-AZn/tip MJs (n=1 to 4, batch #2) :* ***(a-d)*** *pristine MJs after fabrication* ***(e-h)*** *after light irradiation (385-405 nm) to induce the trans-to-cis isomerization and* ***(i-l)*** *after light irradiation (325-365 nm) to induce the cis-to-trans isomerization (except for AZ2, thermal back isomerization). The blue lines are the fits with Gaussian distributions. The mean values and the standard deviations are given in every panel.*

From the SEL model fitting, the large increase in the current is related to a strong variation in the molecule/ electrode coupling energies ($\Gamma_1$ and $\Gamma_2$, see the detail of the SEL model equation in the Supporting Information) between the molecules in the SAM and the electrodes. Figure 6 shows the histograms of the molecule/electrode coupling energies for the 4 samples (batch #2) in the 3 states. The first observation is that the I-V traces are almost symmetric ($\Gamma_1 \approx \Gamma_2$) with values ca. 0.1-1 meV for the pristine $^{TS}$Au-S-AZn/tip MJs and with the azobenzene in the *trans*-state (Fig. 6 a-d and i-l). For the $^{TS}$Au-S-AZn/tip MJs with the molecules in the *cis*-state (Fig. 6 e-h), the coupling energy increases by roughly a decade (≈ 1-10 meV) - Table S5 and Fig. 7.

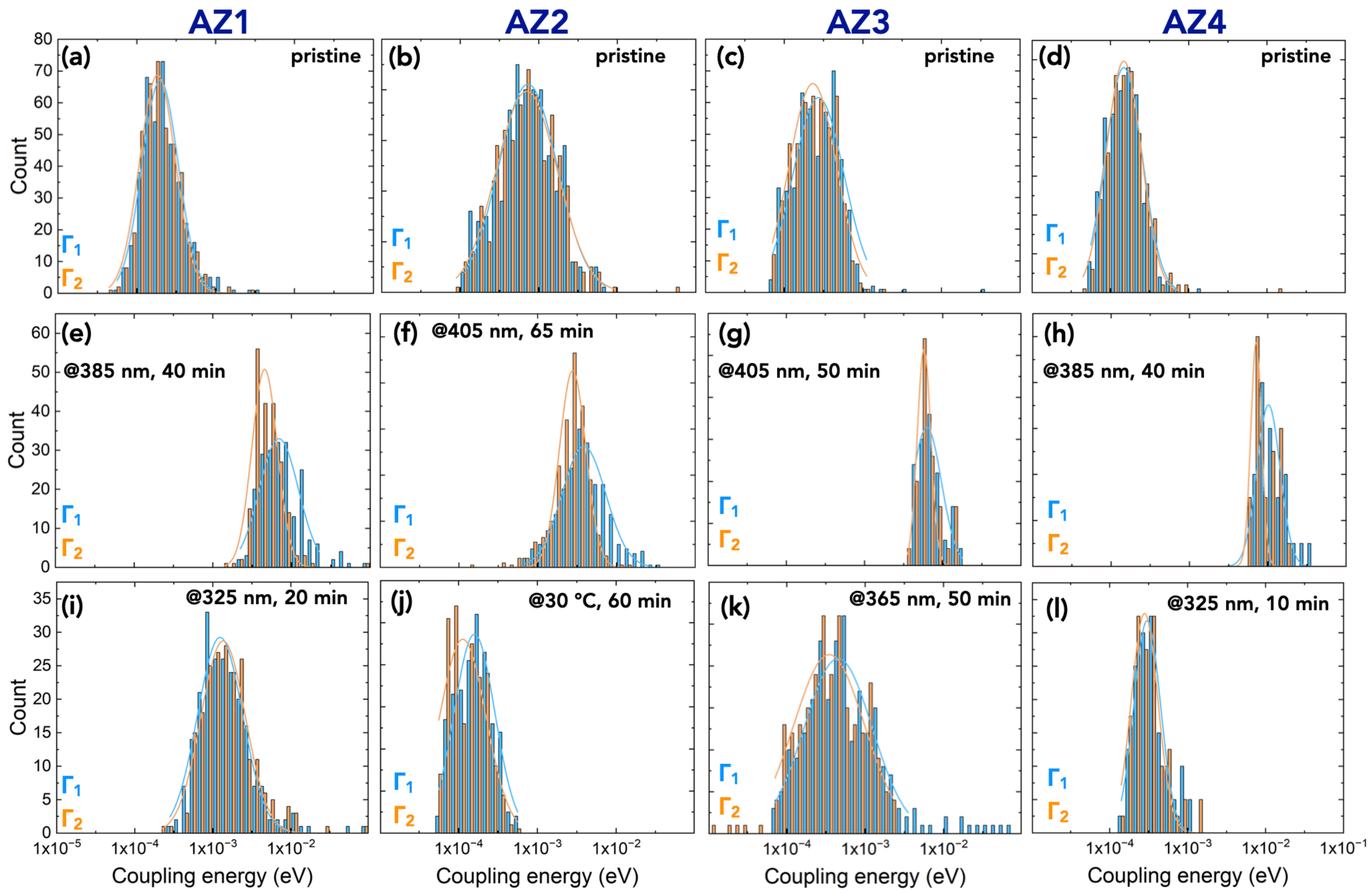

***Figure 6***. *Histograms of the molecule/electrode coupling energies ($\Gamma_1$ and $\Gamma_2$) obtained by fitting all the I-V data sets of the four ${}^{TS}$Au-S-AZn/tip MJs (n=1 to 4, batch #2):* ***(a-d)*** *pristine MJs after fabrication* ***(e-h)*** *after light irradiation (385-405 nm) to induce the trans-to-cis isomerization and* ***(i-l)*** *after light irradiation (325-365 nm) to induce the cis-to-trans isomerization (except for AZ2, thermal back isomerization). The solid lines are the fits with a log-normal distribution (see values Table S5 in the Supporting Information).*

For the samples of batch #1, the same trends are observed for the SEL model parameters (Fig. 7, and Figs. S9-S10). For the two batches, the behavior of the energetics is similar: the HOMO energy level is not significantly affected by the nature of the spacer, nor by the conformation of the azobenzene.  For the samples of batch #1, in the pristine state, $\varepsilon_0$ is in the range of 0.68-0.9 eV, slightly increasing to 0.73-1 eV upon *cis*-isomerization. These values are close to those measured for the samples of batch #2 (*vide supra*). The lower conductances in the *cis*-state of the samples of batch #1 are reflected by lower values of the molecule/electrode coupling energy (Fig. 7, data in Table S6).

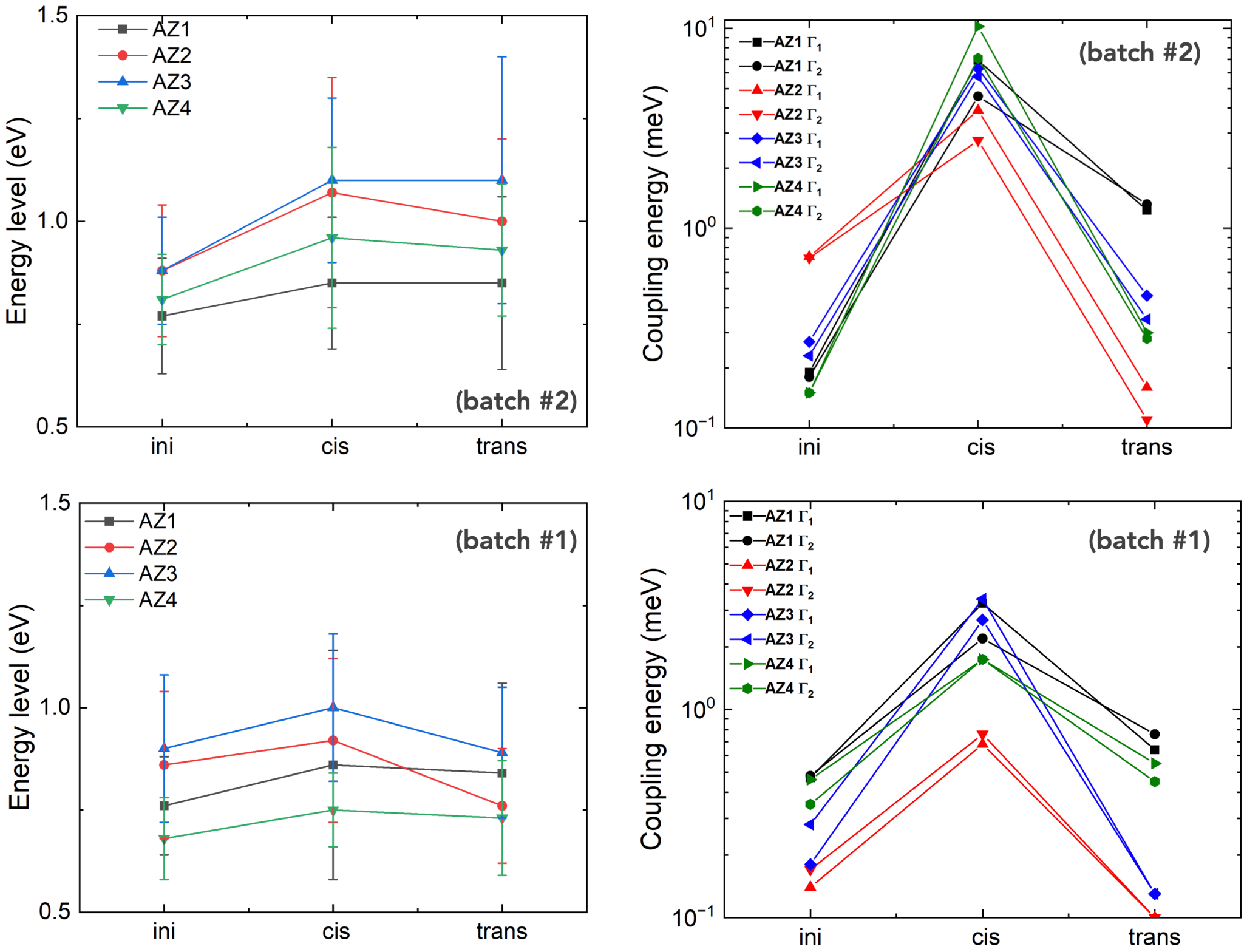

***Figure 7****. Energy level of the HOMO (with respect to the Fermi energy of the electrodes, mean values ± standard deviation, from Fig. 5 for batch #2 and Fig. S9 for batch #1) and molecule/electrode coupling energy (mean values, from Fig. 6 for batch #2 and Fig. S10 for batch #1).*

## 3. THEORETICAL RESULTS.

The geometric structure of the molecules was first optimized in the gas phase at the DFT level (see details in the Supporting Information). The molecules were then anchored on the gold surface via a sulfur atom with a tilt angle chosen to fit the measured SAM thickness (given in Table S2). The unit cell of the gold surface is modeled by a slab of six layers with 2 × 3 Au atoms in each layer and lattice parameters: a = 5.76 Å, b = 8.65 Å and α = 120°. With one molecule per unit cell, this corresponds to a theoretical density of 0.4 $nm^2$/molecule, which fits the range deduced from the ellipsometry measurements (≈ 0.3-0.6 $nm^2$/molecule, see Table S2). The geometry of the interfaces was then optimized by relaxing the molecule and the top gold electrode. The energy-dependent electron transmission function T(ε) is calculated by DFT combined with NEGF as detailed in the Supporting Information. Figure 8 shows the optimized geometries in the *trans* and *cis* states for the AZ1, AZ2 and AZ3 MJs, as well as the calculated energy-dependent transmission coefficients (the case of AZ4 is discussed at the end of this section). For the three MJs in the *trans*-state, the molecules expose the edge of the benzene rings to the top electrodes (Fig. 8, right panels), while in the *cis*-state, the top benzene ring is almost parallel to the surface, thus, the tip is mostly in contact with the benzene π orbitals (Fig. 8, left panels). This feature explains the measured higher conductance of these three MJs when the molecule lies in the *cis*-state in agreement with the calculated higher transmission coefficients at the Fermi energy leading to $T(\varepsilon_F)_{cis}$/

$T(\varepsilon_F)_{trans}$ > 1 (Fig. 8, central panels). Albeit the values of the calculated *cis*/*trans* ratios are lower than the actually measured ones (Fig. 3-b), the trends are qualitatively consistent, with the lowest values for the AZ2 MJ (Fig. 3-b). The quantitative discrepancies between the experimental and theoretical predictions are usually attributed to the fact that cooperative effects (intermolecular interactions, substrate-mediated coupling), which strongly modify the conductance of the ensemble of molecules in the MJs[25-27] are not accurately enough taken into account in single-molecule-like calculations (even with a small unit cell) and that common approximations made in DFT-GGA give imprecise conductance values, although the trends are correctly predicted.[28] The calculated HOMO position is not strongly modified by the isomerization in AZ1 and AZ2 MJs (difference of 0.14-0.2 eV, Table 3), in full consistency with the experiments (Fig. 7); the larger difference calculated for the AZ3 MJ (0.42 eV, Table 3) and not reproduced by the SEL analysis might point to its limitation in large molecules due to the involvement of several orbitals as conduction pathways. Moreover, the largest electrode coupling energies (Fig. 7), reflecting a stronger molecule/electrode coupling, are consistent with the changes in the energy width of the calculated HOMO (Fig. 8, central panels) showing narrow HOMO peaks for the three MJs in the *trans*-state and broader HOMO peaks for the *cis* conformation. The case of the AZ4 MJ is more complicated because the ease of rotation around the triple carbon-carbon bond induces a multitude of conformations in the junction; however, they relax to an almost "planar" geometry for the *trans*-state (*i.e.*, the thiophene and benzene rings facing the substrate) and yield calculated T(ε) curves almost similar for the two isomers (Fig. S12, Supporting Information) and consequently a *cis*/*trans* ratio ≈ 1 (or slightly <1), in contrast to the experimental results (Fig. 3-b). Previous simulations have also clearly reported that the *cis*/*trans* conductance ratios are extremely sensitive to the atomic details at the molecule/top electrode interface (*e.g.*, with values ranging between ≈ 1 and 77).[29] Moreover, the experimental behavior of the AZ4 MJ is not fully reproducible, given the highest *cis*/*trans* ratio of the series for batch #2 and almost the lowest value of the series for batch #1 (Fig. 3-b). We assume that this feature also reflects the great variability of the conformations of the *trans*-AZ4 molecules in the SAMs, leading to the highest value when the azobenzene moiety is oriented similarly to the AZ1-AZ3 counterparts.

| | $HOMO_{cis}$ (eV) | $HOMO_{trans}$ (eV) | $T(\varepsilon_F)_{cis}$ | $T(\varepsilon_F)_{trans}$ | $R_{cis/trans}$ | HOMO-LUMO $gap_{cis}$ (eV) | HOMO-LUMO $gap_{trans}$ (eV) |
|---|---|---|---|---|---|---|---|
| **AZ1** | -0.83 | -0.69 | $1.48 \times 10^{-3}$ | $2.99 \times 10^{-4}$ | 5 | 1.92 | 1.74 |
| **AZ2** | -0.74 | -0.54 | $4.58 \times 10^{-4}$ | $1.45 \times 10^{-4}$ | 3 | 1.89 | 1.87 |
| **AZ3** | -0.87 | -0.45 | $5.00 \times 10^{-4}$ | $2.77 \times 10^{-5}$ | 18 | 1.90 | 1.86 |

***Table 3****. Values of HOMO level energy with respect to the Fermi level have been extracted from the calculated transmission spectra; transmission at $\varepsilon_F$ for cis and trans conformations of AZ1-AZ3 as well as the ratio $R_{cis/trans}$ defined as $T(\varepsilon_F)_{cis}/T(\varepsilon_F)_{trans}$. Calculated HOMO-LUMO gap.*

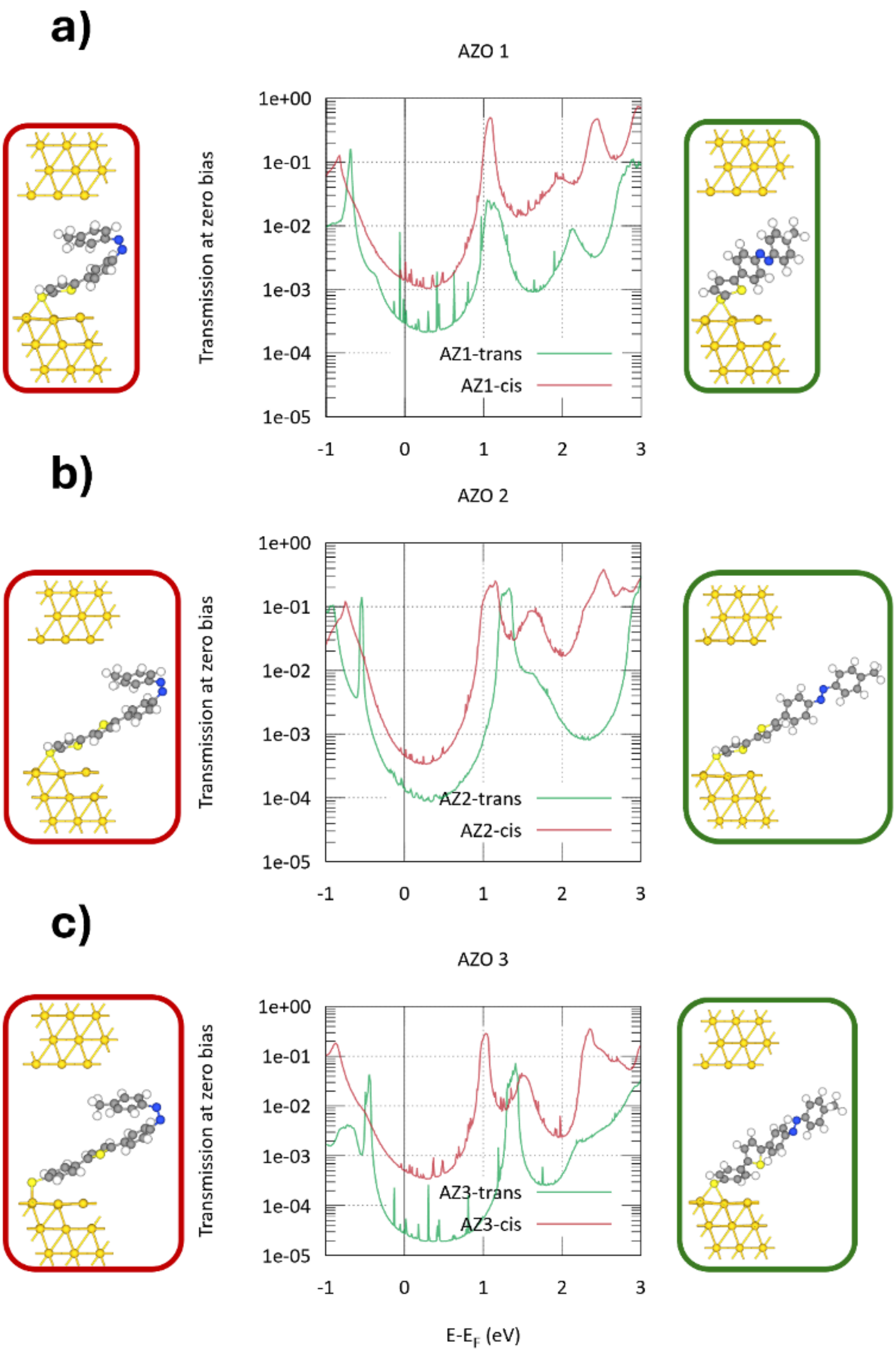

***Figure 8**. Schemes of the optimized atomic structure of the MJs with the molecules in the cis (left panels) and trans (right panels) states and electron transmission functions at zero bias for **(a)** the AZ1 MJ, **(b)** the AZ2 MJ and **(c)** the AZ3 MJ. The cis/trans ratios are calculated at the Fermi energy: $T(\varepsilon_F)_{cis}/T(\varepsilon_F)_{trans}$. Note that the schematic representation is drawn using periodic boundary conditions and includes only the scattering region of the DFT+NEGF formalism for visual clarity.*

## 4. DISCUSSION.

The energy position of the HOMO is experimentally weakly affected by the change in conformation (the variations upon *trans*/*cis* isomerization are smaller than the dispersion of the experimental values). Thus the large variation of the measured current between the two isomers (Fig. 1) cannot be explained by a significant modification of the MO in the MJs. This feature can be correlated with the fact that the HOMO-LUMO gap is not significantly affected by *trans*/*cis* isomerization in our DFT calculations (Fig. 8, Table 3). We note that there is experimentally no direct correlation between the MJ conductances (Fig. 3-a) and the degree of conjugation of the molecules (as assessed by the HOMO-LUMO gap, Table S3). Indeed, for instance, in the *trans*-state, the AZ2 molecule has the smallest HOMO-LUMO gap, but also the smallest conductance (Fig. 3). The electronic properties of the MJs are dominated by the molecular organization in the SAMs and the electronic coupling with the electrodes. The increase in the molecule/electrode energy coupling Γ is the main reason for the increase in conductance for the *cis* $^{TS}$Au-S-AZn/tip MJs. If we assume that the anchoring configuration at the bottom electrode does not change drastically during photoisomerization,[30] the coupling primarily depends on the change in orientation of the top of the molecule. The DFT simulations suggest that the top benzene ring is practically parallel to the top electrode in the *cis* state for AZ1-AZ3 MJs; thus, the coupling is stronger, since the C-AFM tip interacts with the π orbitals rather than with the benzene edge, as also calculated for the simplest azobenzene (no spacer).[29] Surprisingly, both the coupling at the bottom and top electrodes increase together (the I-V curves remain symmetric with respect to the voltage polarity), which can be related to a recently reported interface feedback effect, i.e., that Γ at one interface is not only dependent on the coupling strength at that interface, but also at the second metal interface.[31] In other words, the two molecule/electrode coupling energies are connected through the delocalized frontier orbital of the fully conjugated molecule. Notice that the value of Γ also depends on the estimated number of molecules in the MJ contacted by the C-AFM tip. However, this uncertainty is negligible compared to the large variations (Fig. 7) induced by the trans/cis isomerization of the azobenzene in the MJs (see section IV.4 in the Supporting Information).

In terms of device performances, the results are summarized as follows:

- The best *cis*/*trans* conductance ratios are obtained for the AZ1 and AZ3 MJs (i.e., thiophene and phenylthiophene spacer, respectively). The AZ2 MJ (bithiophene spacer) has the worst *cis*/*trans* conductance ratio (Fig 3-b). Albeit the AZ4 MJ (with an acetylene thiophene spacer) can display a high *cis*/*trans* ratio, this feature is experimentally sample-dependent and theoretically ascribed to a large number of possible molecule conformations due to the facile rotation around the carbon-carbon triple bond.
- The AZ1, AZ3 and AZ4 MJs also have the same efficient *trans*-to-*cis* switching, while the worst is obtained with the AZ2 MJ (Fig. 4-b). We note that the AZ2 SAM is the most closely packed (batch #2, an area of ≈ 0.3 $nm^2$ per molecule vs. 0.4-0.5 for the three other samples, Table S2 in the supporting information). This may contribute to the slower *trans*-to-*cis* kinetics.
- The AZ3 MJ has the slowest *cis*-to-*trans* back isomerization. The back *cis*-to-*trans* isomerization is the most efficient for the AZ4 MJ (Fig. 4-c).

- AZ2 MJ is the worst device with the lowest *cis*/*trans* ratio and the slowest *trans*-to-*cis* light isomerization. Moreover, the AZ2 MJ cannot be switched back to the *trans*-state by light. The absence of an optical *cis*-to-*trans* transition would be explained if the two isomers have almost the same maximum optical absorption wavelength. Thus, the composition of the photo-stationary state is driven by a competition between the two isomerization transition efficiencies (the molar absorption coefficients are almost similar at the irradiation wavelength used in the experiments) and the thermal stability of the *cis* form. This feature may also explain the slow *trans*-to-*cis* switching of the AZ2 MJs. This hypothesis would require more detailed photochemical studies.

## V. CONCLUSIONS.

In conclusion, the MJs with the thiophene (AZ1) and phenylthiophene (AZ3) spacers present the best switching performances (the highest *cis*/*trans* conductance ratios, the fastest *trans*-to-*cis* isomerization switching), at the stated conditions (light wavelength and intensity) indicated in this report and for one batch of samples (#2) out of two. The MJs with a bithiophene spacer (AZ2) show poor behavior (lower *cis*/*trans* conductance ratio and longer *trans*-to-*cis* switching time). This trends are qualitatively consistent with theoretical simulations. Introducing a spacer with a carbon-carbon triple bond (acetylene thiophene, AZ4) leads to irreproducible and unpredictable performances (batch-to-batch inconsistency). In addition, the AZ1 MJ displays a faster *cis*-to-*trans* back isomerization than AZ3. The electrical conductance properties of these azobenzene-based MJs are mainly controlled by the conformation-dependent coupling energy at the molecule/electrode interface, rather than by the energy position of the HOMO level, which is weakly dependent on the nature of the π-conjugated spacer, nor on the conformational states (*trans* or *cis*) of the azobenzene moiety.

## ASSOCIATED CONTENT.

### Data Availability Statement.

The data that support the findings of this study are available from the corresponding author upon reasonable request.

### Supporting Information.

The Supporting Information is available free of charge at....

Synthesis of the molecules and SAMs, details on the ellipsometry and C-AFM measurements, conditions of the light irradiations, additional data, single-energy level parameters and details on the DFT calculations.

## AUTHOR INFORMATION.

### Corresponding author.

**Dominique Vuillaume** - Institute for Electronics Microelectronics and Nanotechnology (IEMN), CNRS, Av. Poincaré, Villeneuve d'Ascq, France. Orcid : https://orcid.org/0000-0002-3362-1669 ; E-mail : dominique.vuillaume@iemn.fr

### Authors.

**Laurette-Apolline Jerro** - Institute for Electronics Microelectronics and Nanotechnology (IEMN), University of Lille, Av. Poincaré, Villeneuve d'Ascq, France. Orcid : https://orcid.org/0009-0000-0916-988X

**Imen Hnid** - Institute for Electronics Microelectronics and Nanotechnology (IEMN), University of Lille, Av. Poincaré, Villeneuve d'Ascq, France. Orcid : https://orcid.org/0000-0002-4468-5334

**Imane Arbouch** - Laboratory for Chemistry of Novel Materials, University of Mons, Belgium. Orcid : https://orcid.org/0000-0002-4593-0098

**Ali Yassin** - Department of Physical Sciences, School of Arts and Sciences, Lebanese American University, Beirut, Lebanon. Orcid : https://orcid.org/0000-0002-3047-9863

**David Guérin** - Institute for Electronics Microelectronics and Nanotechnology (IEMN), CNRS, Av. Poincaré, Villeneuve d'Ascq, France. Orcid : https://orcid.org/0000-0002-4338-1742

**Alexis Elias** - Laboratory for Chemistry of Novel Materials, University of Mons, Belgium. Orcid : https://orcid.org/0009-0009-6545-0604

**Colin van Dyck** - Theoretical Chemical Physics group, University of Mons, Mons, Belgium. Orcid : https://orcid.org/0000-0003-2853-3821

**Lionel Sanguinet** - MOLTECH-Anjou, CNRS, SFR MATRIX, University of Angers, Angers, France. Orcid : https://orcid.org/0000-0002-4334-9937

**Jérôme Cornil** - Laboratory for Chemistry of Novel Materials, University of Mons, Belgium. Orcid : https://orcid.org/0000-0002-5479-4227

**Stéphane Lenfant** - Institute for Electronics Microelectronics and Nanotechnology (IEMN), CNRS, Av. Poincaré, Villeneuve d'Ascq, France. Orcid : https://orcid.org/0000-0002-6857-8752

**Philippe Blanchard** - MOLTECH-Anjou, CNRS, SFR MATRIX, University of Angers, Angers, France. Orcid : https://orcid.org/0000-0002-9408-8108

## Author Contributions.

I.H., L.J.and D.G. fabricated and characterized the SAMs. A.Y. synthesized the molecules. A.Y., L.S. and P.B. discussed the synthetic procedures and performed the physico-chemical and spectroscopic analysis. I.H. and L.J. did the C-AFM measurements. I.A., A.E., C.v.D. and J.C. conducted the theoretical simulations. I.H., L.J. and D.V. analyzed the I-V datasets. S.L., P.B. and D.V. supervised the project. The manuscript was written by D.V. with the contributions and comments of all the authors. All authors have approved the final version of the manuscript.

## Funding.

Funding from the French ANR, project EVOLMONET (#ANR-20-CE30-0002); Fund for Scientific Research (FRS) of FNRS within the Consortium des Equipements de Calcul Intensif (CECI) under grant 2.5020.11 and by the Walloon Region (Tier-1 supercomputer) under grant 1910247; Lebanese American University-President’s Intramural Research Fund (PIRF) #I0042.

## Note.

The authors declare no competing financial interest.

## ACKNOWLEDGEMENTS.

We acknowledge the technical support from the staff of the IEMM scanning probe microscopy platform. J.C. is an FNRS research director.

## REFERENCES.

(1) Feringa, B. L. *Molecular Switches*; Wiley-VCH Verlag GmbH, 2001.

(2) Russew, M.-M.; Hecht, S. Photoswitches: From Molecules to Materials. *Adv. Mater.* **2010**, *22* (31), 3348–3360.

(3) Jerca, F. A.; Jerca, V. V.; Hoogenboom, R. Advances and opportunities in the exciting world of azobenzenes. *Nat. Rev. Chem.* **2022**, *6* (1), 51–69.

(4) Son, H.; Kwak, S.; Noh, H.; Kim, M.; Noh, D.; Chakraborty, S.; Lee, J.; Cho, Y.; Kim, K.; Eom, T.; et al. Azobenzene's Cross-Scale Optics and Photonics: Molecular Photoswitching, Mesoscopic Material Motions, and Adaptive Devices. *Adv. Mater.* **2026**, *38*, e22702.

(5) Xu, Y.; Tang, Y.; Li, Q. Multistate photoswitches beyond on and off. *Nat. Commun.* **2026**, *17* (1), 4029.

(6) Katsonis, N.; Lubomska, M.; Pollard, M.; Feringa, B. L.; Rudolf, P. Synthetic light-activated molecular switches and motors on surfaces. *Prog. Surf. Sci.* **2007**, *82* (7-8), 407–434.

(7) Ferri, V.; Elbing, M.; Pace, G.; Dickey, M. D.; Zharnikov, M.; Samori, P.; Mayor, M.; Rampi, M. A. Light-powered electrical switch based on cargo-lifting azobenzene monolayers. *Angew. Chem. Int. Ed.* **2008**, *47* (18), 3407–3409.

(8) van der Molen, S. J.; Liljeroth, P. Charge transport through molecular switches. *J Phys Condens Matter* **2010**, *22* (13), 133001.

(9) Yasuda, S.; Nakamura, T.; Matsumoto, M.; Shigekawa, H. Phase Switching of a Single Isomeric Molecule and Associated Characteristic Rectification. *Journal of the American Chemical Society* **2003**, *125* (52), 16430–16433.

(10) Ahonen, P.; Laaksonen, T.; Schiffrin, D. J.; Kontturi, K. Photoswitching electron transport properties of an azobenzene containing thiol-SAM. *Phys. Chem. Chem. Phys.* **2007**, *9*, 4898–4901.

(11) Kumar, A. S.; Ye, T.; Takami, T.; Yu, B.-C.; Flatt, A. K.; Tour, J. M.; Weiss, P. S. Reversible Photo-Switching of Single Azobenzene Molecules in Controlled Nanoscale Environments. *Nano Lett.* **2008**, *8* (6), 1644–1648.

(12) Suda, M.; Kameyama, N.; Ikegami, A.; Einaga, Y. Reversible Phototuning of the Large Anisotropic Magnetization at the Interface between a Self-Assembled Photochromic Monolayer and Gold. *Journal of the American Chemical Society* **2008**, *131* (2), 865–870.

(13) Jung, U.; Filinova, O.; Kuhn, S.; Zargarani, D.; Bornholdt, C.; Herges, R.; Magnussen, O. Photoswitching Behavior of Azobenzene-Containing Alkanethiol Self-Assembled Monolayers on Au Surfaces. *Langmuir* **2010**, *26* (17), 13913–13923.

(14) Ferri, V.; Elbing, M.; Pace, G.; Dickey, M. D.; Zharnikov, M.; Samori, P.; Mayor, M.; Rampi, M. A. Light-powered electrical switch based on cargo-lifting azobenzene monolayers. *Angew Chem Int Ed Engl* **2008**, *47* (18), 3407–3409.

(15) Mativetsky, J. M.; Pace, G.; Elbing, M.; Rampi, M. A.; Mayor, M.; Samori, P. Azobenzenes as Light-Controlled Molecular Electronic Switches in Nanoscale Metal-Molecule-Metal Junctions. *J. Am. Chem. Soc.* **2008**, *130* (29), 9192–9193.

(16) Smaali, K.; Lenfant, S.; Karpe, S.; Oçafrain, M.; Blanchard, P.; Deresmes, D.; Godey, S.; Rochefort, A.; Roncali, J.; Vuillaume, D. High On−Off Conductance Switching Ratio in Optically-Driven Self-Assembled Conjugated Molecular Systems. *ACS Nano* **2010**, *4* (4), 2411–2421.

(17) Comstock, M. J.; Levy, N.; Kirakosin, A.; Cho, J.; Lauterwasser, F.; Harvey, J. H.; Strubbe, D. A.; Fréchet, J. M. J.; Trauner, D.; Louie, S. G.; et al. Reversible photomechanical switching of individual engineered molecules at a metallic surface. *Phys. Rev. Lett.* **2007**, *99*, 038301.

(18) Wu, S. D.; Chen, Z. Z.; Sun, W. J.; Shi, L. Y.; Shen, A. K.; Cao, J. J.; Liu, Z.; Lambert, C. J.; Zhang, H. L. Boosting the Photoresponse of Azobenzene Single-Molecule Junctions via Mechanical Interlock and Dynamic Anchor. *ACS Nano* **2024**, *18* (45), 31547–31558.

(19) Datta, S. *Electronic transport in mesoscopic systems*; Cambridge University Press, 1995.

(20) Cuevas, J. C.; Scheer, E. *Molecular Electronics: An introduction to theory and experiment*; World Scientific, 2010.

(21) Landauer, R. Spatial variation of currents and fields due to localized scatterers in metallic conduction. *IBM Journal* **1957**, (July), 223–231.

(22) Büttiker, M.; Imry, Y.; Landauer, R.; Pinhas, S. Generalized many-channel conductance formula with application to small rings. *Phys. Rev. B* **1985**, *31* (10), 6207–6215.

(23) Yamaguchi, H.; Matsuda, K. Photo- and Electrochromic Switching of Diarylethene–Gold Nanoparticle Network on Interdigitated Electrodes. *Chem. Lett.* **2009**, *38* (10), 946–947.

(24) Niebel, C.; Calard, F.; Jarrosson, T.; Lere-Porte, J. P.; Breton, T.; Serein-Spirau, F. Spontaneous assembly of silylethane-thiol derivatives on Au(111): a chemically robust thiol protecting group as the precursor for the direct formation of aromatic gold thiolate monolayers. *Chem. Commun.* **2015**, *51* (36), 7622–7625.

(25) Magoga, M.; Joachim, C. Conductance of molecular wires connected or bonded in parallel. *Phys. Rev. B* **1999**, *59*, 16011–16021.

(26) Reuter, M. G.; Seideman, T.; Ratner, M. A. Molecular Conduction through Adlayers: Cooperative Effects Can Help or Hamper Electron Transport. *Nano Letters* **2011**, *11* (11), 4693–4696.

(27) Trasobares, J.; Rech, J.; Jonckheere, T.; Martin, T.; Aleveque, O.; Levillain, E.; Diez-Cabanes, V.; Olivier, Y.; Cornil, J.; Nys, J. P.; et al. Estimation of π-π Electronic Couplings from Current Measurements. *Nano Letters* **2017**, *17* (5), 3215–3224.

(28) Darancet, P.; Widawsky, J. R.; Choi, H. J.; Venkataraman, L.; Neaton, J. B. Quantitative Current–Voltage Characteristics in Molecular Junctions from First Principles. *Nano Lett.* **2012**, *12* (12), 6250–6254.

(29) Osella, S.; Samorì, P.; Cornil, J. Photoswitching Azobenzene Derivatives in Single Molecule Junctions: A Theoretical Insight into the I/V Characteristics. *J. Phys. Chem. C* **2014**, *118* (32), 18721–18729.

(30) Pace, G.; Ferri, V.; Grave, C.; Elbing, M.; Von Hanisch, C.; Zharnikov, M.; Mayor, M.; Rampi, M. A.; Samori, P. Cooperative light-induced molecular movements of highly ordered azobenzene self-assembled monolayers. *Proceedings of the National Academy of Sciences* **2007**, *104* (24), 9937–9942.

(31) Feng, Y.; Chen, J.; Bâldea, I.; Frisbie, C. D.; Xie, Z. Interface Feedback Effect in Molecular Tunnel Junctions. *J. Am. Chem. Soc. Au* **2025**, *5* (3), 1258–1267.

## ToC graphic.

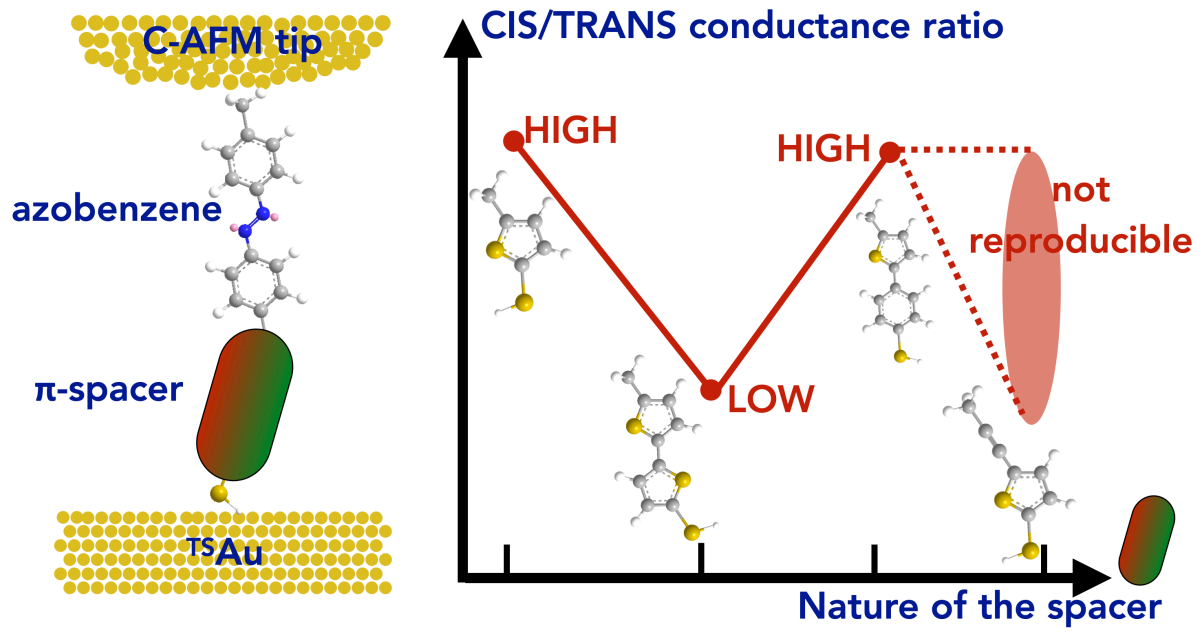

# Modulations of the photoswitching performances of azobenzene-based molecular devices by the π-conjugated spacers.

Laurette-Apolline Jerro,[1] Imen Hnid,[1] Imane Arbouch,[2] Ali Yassin,[3,4] David Guérin,[1]
Alexis Elias,[2] Colin van Dyck,[5] Lionel Sanguinet,[3] Jérôme Cornil,[2] Stéphane Lenfant,[1]
Philippe Blanchard[3] and Dominique Vuillaume.[1*]

*1) Institute for Electronics Microelectronics and Nanotechnology (IEMN), CNRS, University of Lille, Av. Poincaré, Villeneuve d'Ascq, France.*
*2) Laboratory for Chemistry of Novel Materials, University of Mons, Belgium.*
*3) MOLTECH-Anjou, CNRS, SFR MATRIX, University of Angers, Angers, France.*
*4) Department of Physical Sciences, School of Arts and Sciences, Lebanese American University, Beirut, Lebanon.*
*5) Theoretical Chemical Physics group, University of Mons, Mons, Belgium.*

** Corresponding authors: dominique.vuillaume@iemn.fr*

## Supporting Information.

## I. Synthesis and characterizations of the molecules.

***General information.***

Reagents and chemicals from commercial sources were used without further purification. Solvents were freshly dried and purified by distillation from sodium–benzophenone ketyl radical. Flash chromatography was performed with analytical-grade solvents using Aldrich silica gel (technical grade, pore size 60 Å, 230-400 mesh particle size). Flexible plates ALUGRAM® Xtra SIL G UV254 from MACHEREY-NAGEL were used for TLC. Compounds were detected by UV irradiation (Bioblock Scientific). Compounds purified with recycling preparative size exclusion HPLC were solubilized in HPLC grade chloroform. Prior to injection, the solution was filtered through a 0.45 µm PTFE filter (VWR 25 mm syringe filter w/ 0.45 µm membrane). Purification was performed on a LC-9160 NEXT system from the Japan Analytical Industry Co., Ltd. (JAI) equipped with coupled UV-Vis 4Ch NEXT and RI-700 II detectors at room temperature through a set of two JAIGEL-2H and 2.5H columns at an elution rate of 10 mL.min$^{-1}$. NMR spectra were recorded with a Bruker AVANCE III 300 ($^{1}$H, 300 MHz and $^{13}$C, 75 MHz) or a Bruker AVANCE DRX500 ($^{1}$H, 500 MHz; $^{13}$C, 125 MHz). Chemical shifts are given in ppm relative to TMS and coupling constant J in Hz. UV-Vis spectra were recorded with a Perkin Elmer 950 spectrometer. Matrix Assisted Laser Desorption/Ionization was performed on MALDI-TOF MS BIFLEX III Bruker Daltonics spectrometer. High-resolution mass spectrometry (HRMS) was performed with a JEOL JMS-700 B/E or a JEOL Spiral-TOF JMS3000.

***Organic synthesis and characterizations.***

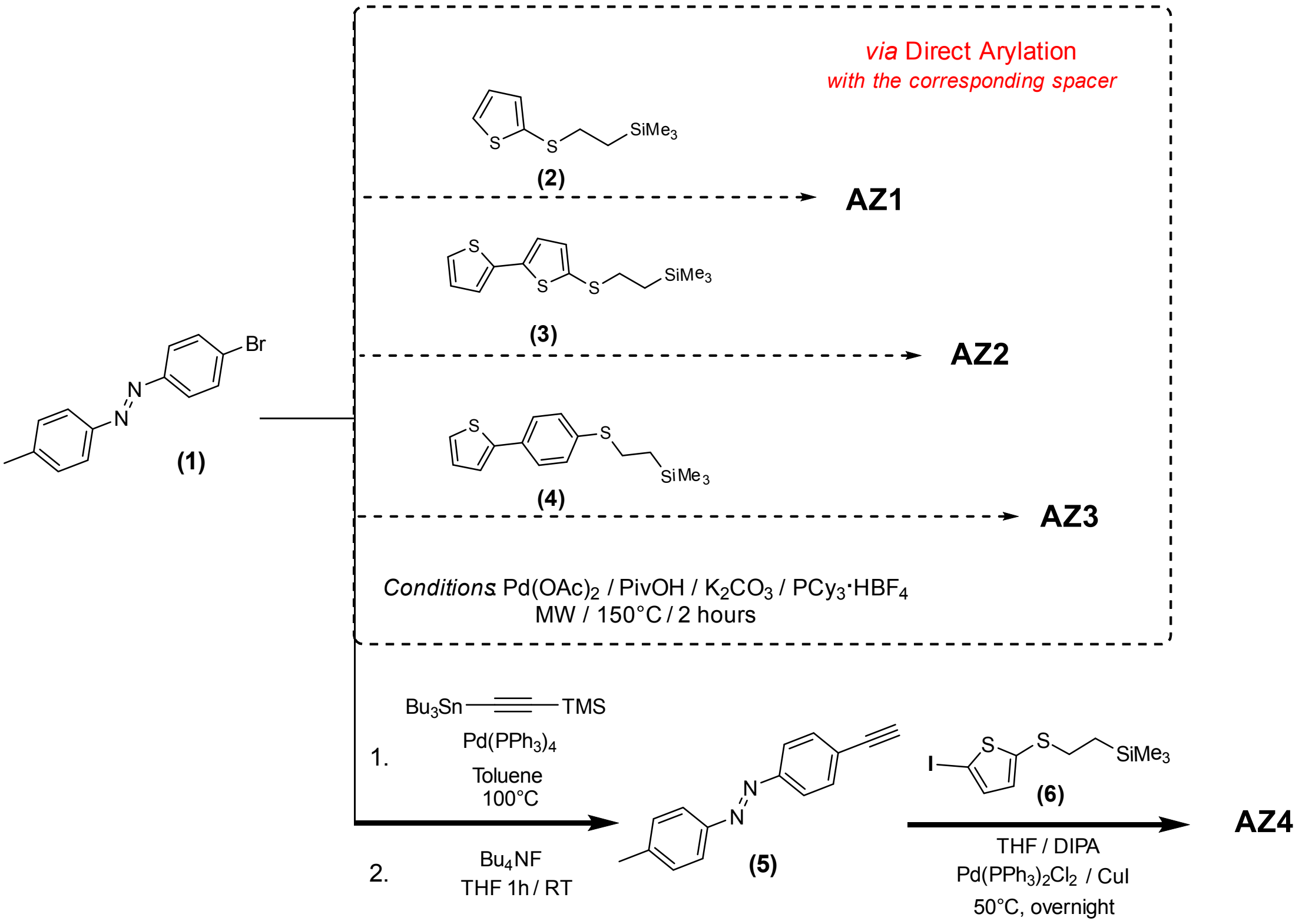

Compounds AZ1, AZ2, and AZ3 were synthesized according to an optimized Pd-catalyzed C–H direct arylation method. 4-Bromo-4'-methyl-azobenzene **(1)**,[1] trimethyl(2-(thiophen-2-ylthio)ethyl)silane **(2)**,[2] (2-([2,2'-bithiophen]-5-ylthio)ethyl)trimethylsilane **(3)** and trimethyl(2-((4-(thiophen-2-yl)phenyl)thio)ethyl)silane **(4)**,[3] were synthesized according to reported procedures.

General procedure for the synthesis of azobenzene derivatives **AZ1–AZ3**: To an oven-dried microwave tube containing a stirring bar, 4-Bromo-4'-methyl-azobenzene **(1)** (100 mg, 0.36 mmol) and the corresponding aryl derivative (**2**, **3**, or **4)** (1 eq, 0.36 mmol), palladium acetate $Pd(OAc)_2$ (0.1 eq, 0.03 mmol), tricyclohexylphosphine tetrafluoroborate $PCy_3{\cdot}HBF_4$ (0.2 eq, 0.07 mmol), pivalic acid PivOH (1 eq, 0.36 mmol) and potassium carbonate (3 eq, 1.1 mmol) were sequentially added before a septum cap was crimped on the vial to form a seal. The solids were degassed 3 times by vacuum-argon filling cycles. Dry toluene (4 mL) was then added to the reaction mixture before exposing to microwave irradiation for 2 hours at 150 °C. The resulting mixture was cooled to room temperature, concentrated under reduced pressure and the residue was purified by column chromatography on silica gel (eluent:

petroleum ether/DCM with 4/1 ratio) affording the corresponding product as a shiny orange solid with yields ranging 30–60%.

**1-(p-tolyl)-2-(4-(5-((2-(trimethylsilyl)ethyl)thio)thiophen-2-yl)phenyl)diazene (AZ1)**:

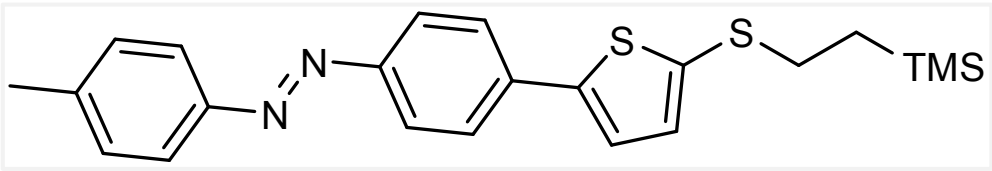

Following the procedure described above gives 82 mg (55% yield) of a shiny orange solid. **$^{1}$H NMR** (300 MHz, $CDCl_3$) δ 9.86 (d, J = 8.6 Hz, 2H), 9.77 (d, J = 8.3 Hz, 2H), 9.66 (d, J = 8.6 Hz, 2H), 9.32 – 9.25 (m, 2H), 4.95 – 4.77 (m, 1H), 4.38 (s, 3H), 3.47 (s, 4H), 2.91 (d, J = 17.7 Hz, 3H), 1.97 (s, 9H). **$^{13}$C NMR** (125 MHz, $CD_2Cl_2$): δ 152.18, 151.23, 151.12, 146.33, 142.35, 137.09, 136.82, 136.63, 134.37, 130.20, 126.39, 124.76, 123.86, 123.54, 123.19, 123.05, 35.52, 21.64, 17.94. **HRMS** (EI) m/z: Calc for $C_{22}H_{26}N_2S_2Si$ ($M^+$), 410.1307; Found, 410.1323.

**1-(p-tolyl)-2-(4-(5'-((2-(trimethylsilyl)ethyl)thio)-[2,2'-bithiophen]-5-yl)phenyl)diazene (AZ2)**:

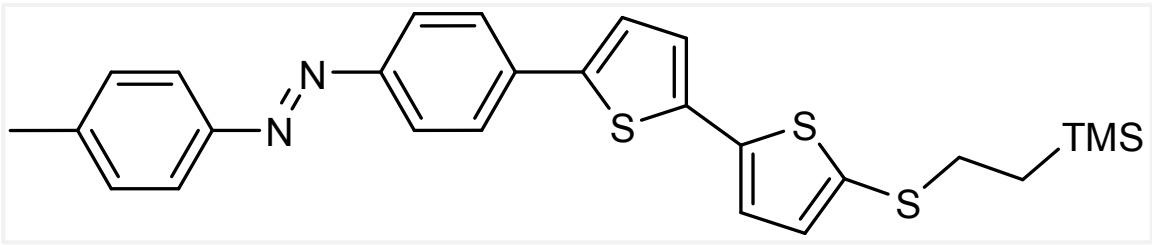

Following the procedure described above gives 42 mg (30% yield) of a shiny orange solid. **$^{1}$H NMR** (500 MHz, $CD_2Cl_2$): δ 7.88 (d, J = 1.8 Hz, 2H), 7.86 (d, J = 2.0 Hz, 2H), 7.77 (d, J = 1.8 Hz, 2H), 7.75 (d, J = 1.9 Hz, 2H), 7.72 (d, J = 2.0 Hz, 2H), 7.70 (d, J = 1.9 Hz, 2H), 7.53 (s, 1H), 7.50 (s, 1H), 7.37 (s, 1H), 7.36 (s, 1H), 7.30 – 7.27 (m, 2H), 7.26 (d, J = 0.8 Hz, 2H), 2.37 (s, 4H), 1.45 (s, 1H), 0.95 – 0.83 (m, 2H), 0.08 (s, 1H). **$^{13}$C NMR** (125 MHz, $CD_2Cl_2$): δ 152.23, 151.28, 142.60, 142.33, 140.29, 137.99, 137.83, 136.54, 135.56, 134.22, 130.19, 126.39, 125.32, 125.28, 124.45, 123.88, 123.19, 35.60, 30.09, 17.96. **HRMS** (EI) m/z: Calc for $C_{26}H_{28}N_2S_3Si$ ($M^+$), 492.1184; Found, 492.1127.

**1-(p-tolyl)-2-(4-(5-(4-((2-(trimethylsilyl)ethyl)thio)phenyl)thiophen-2-yl)phenyl)diazene (AZ3)**:

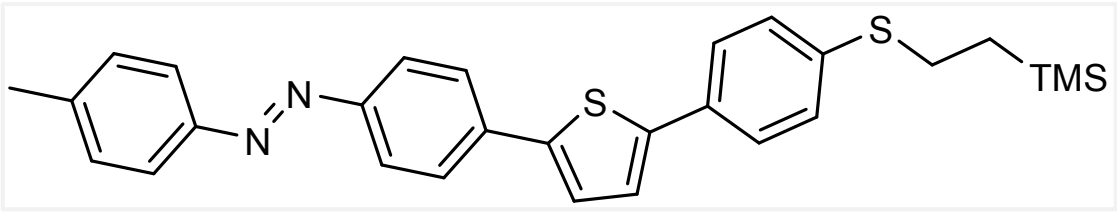

Following the procedure described above gives 105 mg (60% yield) of a shiny orange solid. **$^{1}$H NMR** (300 MHz, $CDCl_3$) δ 7.94 (d, J = 8.4 Hz, 2H), 7.85 (d, J = 7.6 Hz, 2H), 7.76 (d, J = 7.9 Hz, 2H), 7.59 – 7.54 (m, 2H), 7.40 (d, J = 3.8 Hz, 2H), 7.35 – 7.30 (m, 2H), 3.06 – 2.92 (t, 3H), 2.45 (s, 2H), 1.55 (s, 3H), 1.01 – 0.90 (m, 2H), 0.16 (d, J = 3.2 Hz, 9H). **$^{13}$C NMR** (125 MHz, $CD_2Cl_2$): δ 152.10, 151.25, 144.65, 142.59, 142.31, 137.89, 136.86, 131.63, 130.20, 129.15, 126.33, 126.26, 125.64, 125.60, 124.52, 123.87, 123.24, 123.18, 29.57, 21.64, 17.16. **HRMS** (EI) m/z: Calc for $C_{28}H_{30}N_2S_2Si$ ($M^+$), 486.1620; Found, 486.1677.

**1-(*p*-tolyl)-2-(4-((trimethylsilyl)ethynyl)phenyl)diazene:**

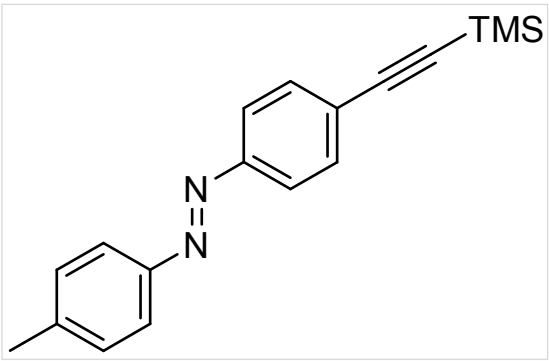

Trimethyl((tributylstannyl)ethynyl)silane (250 mg, 0.65 mmol) was added to a solution of 4-bromo-4'-methyl-azobenzene (**1**) (155 mg, 0.56 mmol) and $Pd(PPh_3)_4$ (39 mg, 6 mol%) in 15 mL of freshly distilled dry toluene under Argon atmosphere. The mixture was stirred under an inert atmosphere at 100°C overnight, before being cooled to room temperature. The reaction was quenched with water, extraction was performed three times with dichloromethane, and the combined organic layers were dried over $MgSO_4$. After the removal of the solvent, the residue was purified by chromatography on silica gel (20% dichloromethane in petroleum ether as eluent) to give the product as yellow powder (160 mg, 97% yield). **$^1$H NMR** (300 MHz, $CDCl_3$): δ 7.83 (m, 4H), 7.59 (m, 2H), 7.31 (d, J = 8.2, 2H), 2.44 (s, 3H), 0.27 (s, 9H). **HRMS** (EI) m/z: Calc for $C_{18}H_{20}N_2Si$ ($M^+$), 292.1396; Found, 292.1358.

**1-(4-ethynylphenyl)-2-(p-tolyl)diazene (5):**

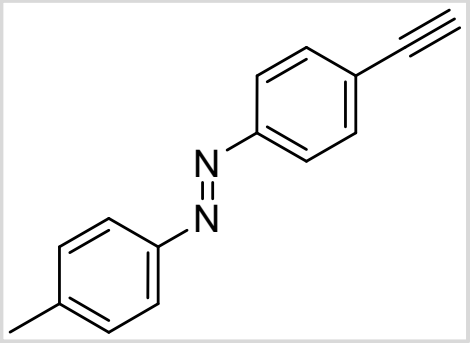

To a solution of 1-(*p*-tolyl)-2-(4-((trimethylsilyl)ethynyl)phenyl)diazene (160 mg, 0.55 mmol) in 10 mL of THF, a 1M solution of tetra-n-butylammonium fluoride was carefully added dropwise (0.6 mL, 157 mg, 0.6 mmol), after which the reaction was stirred at room temperature for 1 hour. The solvent was evaporated and the mixture was diluted with dichloromethane and washed 3 times with water. The organic layer was dried over anhydrous $Na_2SO_4$ and evaporated in vacuo. The crude compound was quickly flashed through a short chromatography column (silica gel; eluent: dichloromethane) to give 120 mg (99% yield) of the desired yellow solid, which was directly engaged in the next step. **$^1$H NMR** (300 MHz, $CDCl_3$): δ 7.84 (m, 4H), 7.62 (m, 2H), 7.31 (d, J = 8.4, 2H), 3.21 (s, 1H), 2.44 (s, 3H). **HRMS** (EI) m/z: Calc for $C_{15}H_{12}N_2$ ($M^+$), 220.1000; Found, 220.1078.

**(2-bromoethyl)trimethylsilane**:

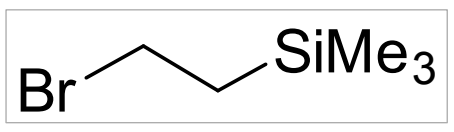

Under argon atmosphere, a solution of phosphorus tribromide (0.4 mL, 2.2 mmol) in anhydrous $CH_2Cl_2$ was added dropwise to a mixture of vinyltrimethylsilane (1 mL, 6.8 mmol) and silica gel (3 g) in anhydrous $CH_2Cl_2$ cooled to -10°C. After 10 min of stirring at this temperature, the reaction mixture was allowed to warm to 20°C over a period of 0.5 h, after which silica gel was separated by filtration. The filtrated solution was washed with a saturated aqueous solution of $Na_2CO_3$, dried over $Na_2SO_4$, and concentrated to dryness to give pure (2-bromoethyl)trimethylsilane as slightly yellow oil (600 mg, 50% yield). The compound is not stable at room temperature, even when stored at +4°C, hence was directly used after synthesis. **$^1$H NMR** (300 MHz, $CDCl_3$): δ 3.59–3.55 (m, 2H), 1.39–1.36 (m, 2H), 0.04 ppm (s, 9H).

**2-((5-iodothiophen-2-yl)thio)ethyl trimethylsilane (6)**:

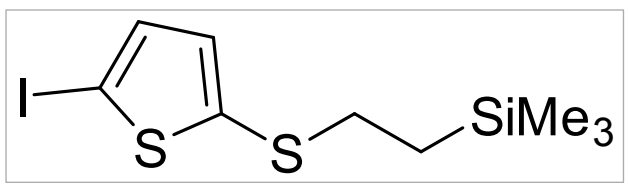

Under Argon atmosphere, a solution of $CsOH.H_2O$ (205 mg, 1.22 mmol) in dry MeOH (2 mL) was added dropwise to a solution of 3-((5-iodothiophen-2-yl)thio)propanenitrile (**4**) (300 mg, 1.02 mmol) in 5 mL of dry DMF. Compound 4 was prepared according to a procedure previously reported by us.[4] The reaction mixture was stirred for 1 h at room temperature before the addition of a solution of (2-bromoethyl)trimethylsilane (221 mg, 1.22 mmol) in 2 mL dry DMF. After 4 hours of additional stirring at 20°C and evaporation of the solvents, the residue was dissolved in $CH_2Cl_2$ and the organic phase was washed with water, dried over $MgSO_4$, and concentrated under reduced pressure. Purification by chromatography on silica gel (100% petroleum ether as eluent) gave the product as dark yellow oil (230 mg, 66% yield). $^1$H NMR (500 MHz, $CDCl_3$): δ 7.12 (d, J = 3.7, 1H), 6.78 (d, J = 3.7, 1H), 2.82 (t, 2H), 0.89 (t, 2H), 0.01 (s, 9H). **$^{13}$C NMR** (125 MHz,

$CDCl_3$): δ 141.1, 137.9, 135.3, 75.7, 35.9, 18.0, -1.3. **HRMS** (EI) m/z: Calc for $C_9H_{15}IS_2Si$ ($M^+$), 341.9429; Found, 341.9387.

**1-(p-tolyl)-2-(4-((5-((2-(trimethylsilyl)ethyl)thio)thiophen-2-yl)ethynyl)phenyl)diazene (AZ4)**:

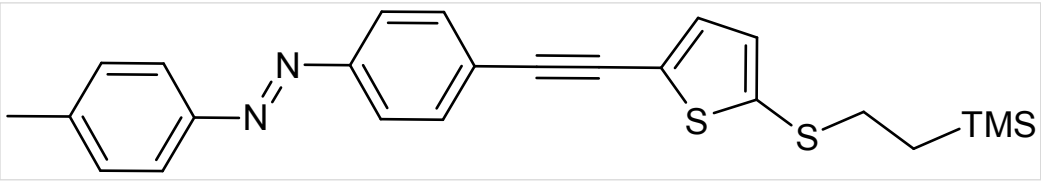

A mixture of 1-(4-ethynylphenyl)-2-(p-tolyl)diazene (**2**) (120 mg, 0.55 mmol), 2-((5-iodothiophen-2-yl)thio)ethyltrimethylsilane (**3**) (240 mg, 0.70 mmol), bis(triphenylphosphine) palladium(II) dichloride (27 mg, 6 mol%), and copper(I) iodide (7 mg, 0.04 mmol) was degassed several times under argon atmosphere, before adding freshly distilled diisopropylamine (5 mL) and THF (5 mL). The mixture was heated under an inert atmosphere at 50°C overnight, after which the solvent was removed by rotary evaporation and the residue was diluted with water and extracted with dichloromethane. The combined organic layers were dried over anhydrous $Na_2SO_4$, evaporated in vacuo and the product was flashed through a short plug of silica to remove the catalyst and baseline impurities. Then, to reach a high degree of purity, the resulting product was injected in a recycling preparative HPLC following the procedure described above to give 85 mg (0.2 mmol, 36% yield) of a shiny orange solid. **$^1$H NMR** (500 MHz, Dichloromethane-$d_2$): δ 7.65 (d, J = 8.4, 2H), 7.56 (d, J = 8.2, 2H), 7.43 (d, J = 8.4, 2H), 7.11 (d, J = 8.2, 2H), 7.01 (d, J = 3.7, 1H), 6.81 (d, J = 3.6, 1H), 2.70 (t, 2H) 2.14 (s, 3H), 0.67 (t, 2H), -0.25 (s, 9H). **$^{13}$C NMR** (125 MHz, Acetone-$d_6$): δ 152.90, 151.62, 143.25, 139.52, 134.17, 133.56, 133.26, 133.11, 130.80, 125.92, 125.53, 123.78, 94.09, 85.24, 35.24, 21.46, 18.15. **HRMS** (EI) m/z: Calc for $C_{24}H_{26}N_2S_2Si$ ($M^+$), 434.1307; Found, 434.1326.

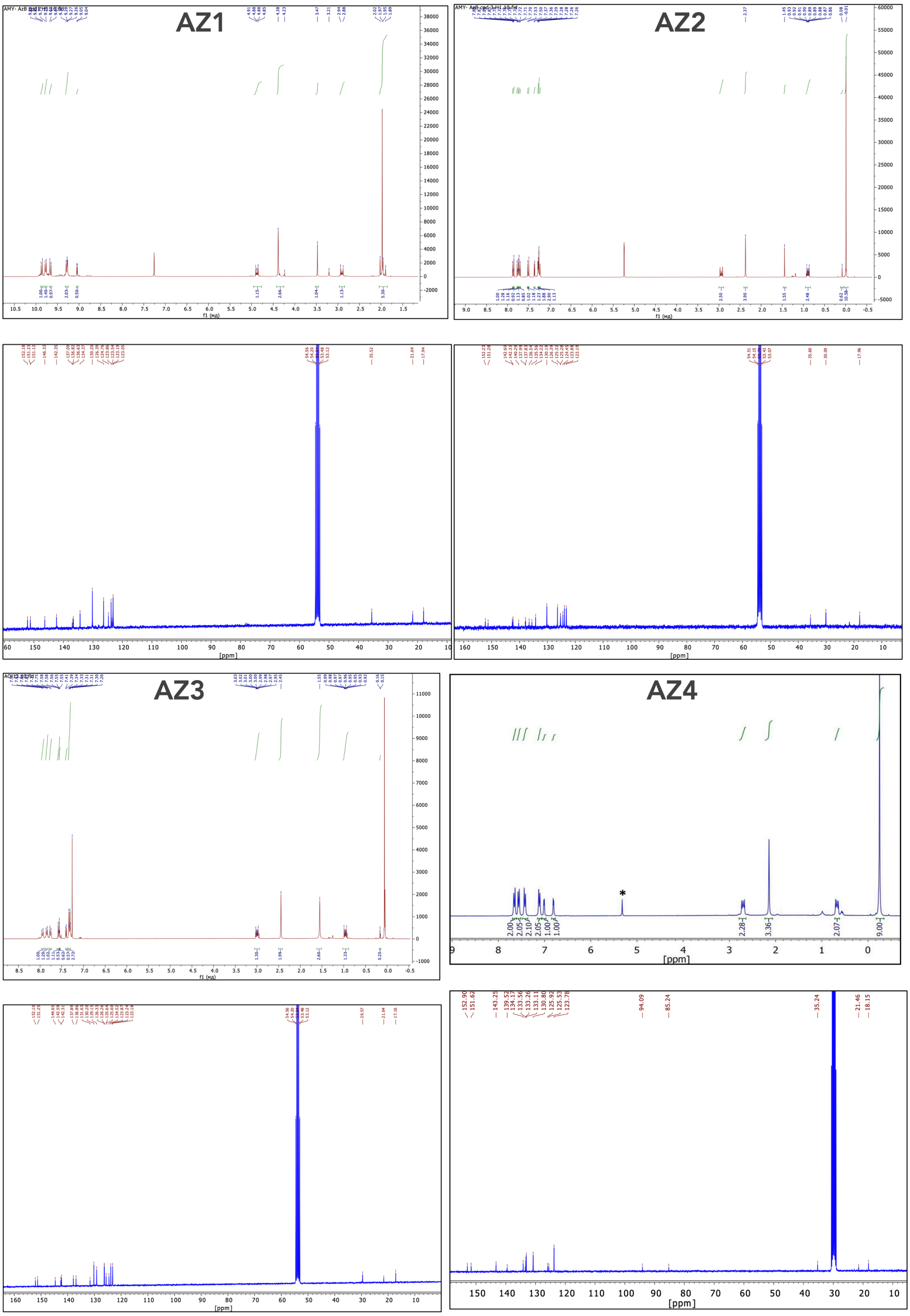

***Figure S1.*** *$^{1}H$ NMR and $^{13}C$ NMR spectra of AZ1, AZ2, AZ3, and AZ4 respectively, recorded at 20°C in $CD_2Cl_2$, $CDCl_3$, or $(CD_3)_2CO$.*

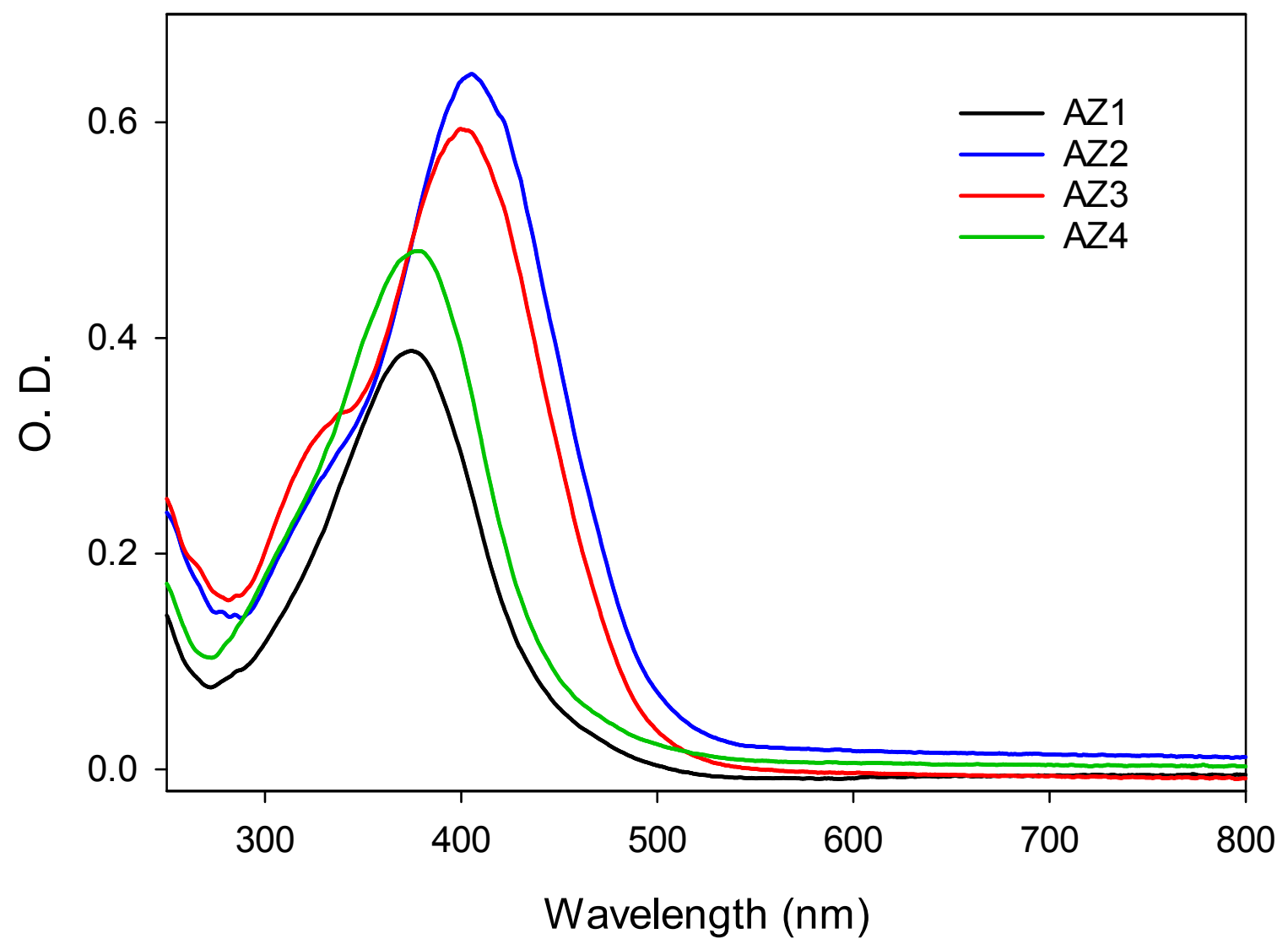

***Figure S2***. *UV-Vis absorption spectra of AZ1, AZ2, AZ3, and AZ4 in the trans-state, recorded at RT in DCM solutions.*

Table S1 reports the wavelength of the maximum of optical absorption and the optical HOMO-LUMO gap measured by optical absorption at the low energy onset of the four molecules in dichloromethane solutions (*trans* state).

| | $\lambda_{max}$(nm) | $E_G$(eV) |
|---|---|---|
| **AZ1** | 373 | 2.38 |
| **AZ2** | 405 | 2.27 |
| **AZ3** | 401 | 2.35 |
| **AZ4** | 378 | 2.42 |

***Table S1***. *Wavelength at the maximum of optical absorption and HOMO-LUMO gap determined at the low energy onset of the spectra (Fig. S2).*

## II. Self-assembled monolayers on Au electrodes.

Ultraflat template-stripped gold surfaces ($^{TS}$Au), with rms roughness of ≈ 0.4 nm were prepared according to methods already reported.[5-7] In brief, a 300−500 nm thick Au film was evaporated on a very flat silicon wafer covered by its native $SiO_2$ (rms roughness of ≈ 0.4 nm), which was previously carefully cleaned by piranha solution (30 min in 7:3 H2SO$_4$/$H_2O_2$ (v/v); **Caution**: Piranha solution is a strong oxidizer and reacts exothermically with organics), rinsed with deionized (DI) water, and dried under a stream of nitrogen. Clean 10x10 mm pieces of glass slide (ultrasonicated in acetone for 5 min, ultrasonicated in 2-propanol for 5 min, and UV irradiated in ozone for 10 min) were glued to the evaporated Au film (UV-polymerizable glue, NOA61 from Epotecny), then mechanically peeled off providing the $^{TS}$Au film attached on the glass side (Au film is cut with a razor blade around the glass piece).

The self-assembled monolayers (SAMs) of the AZn molecules were prepared from a 1 mM solution of the molecules in ethanol/THF (50:50). The $^{TS}$Au substrates were immersed for 16 hours in this solution, and then cleaned in THF with ultrasound for one minute. Two batches were fabricated and characterized.

## III. Ellipsometry.

We recorded spectroscopic ellipsometry data (on ca. 1 $cm^2$ samples) in the visible range using a UVISEL (Horiba Jobin Yvon) spectroscopic ellipsometer equipped with DeltaPsi 2 data analysis software. The system acquired a spectrum ranging from 2 to 4.5 eV (corresponding to 300−750 nm) with intervals of 0.1 eV (or 15 nm). The data were taken at an angle of incidence of 70°, and the compensator was set at 45°. We fit the data by a regression analysis to a film-on-substrate model as described by their thickness and their complex refractive indexes. First, a background for the substrate before monolayer deposition was recorded. We acquired three reference spectra at three different places on the surface spaced a few mm. Secondly, after the monolayer deposition, we acquired once again three spectra at three different places on the surface and we used a 2-layer model (substrate/SAM) to fit the measured data and to determine the SAM thickness. We employed the previously measured optical properties of the substrate (background), and we fixed the refractive index of the monolayer at 1.50.[8] We note that a change from 1.50 to 1.55 would result in less than a 1 Å error for a thickness of less than 30 Å. The three spectra measured on the sample were fitted separately using each of the three reference spectra, giving nine values for the SAM thickness. We calculated the mean value from these nine thickness values and the thickness incertitude corresponding to the standard deviation. Overall, we estimated the accuracy of the SAM thickness measurements at ± 2 Å.[9]

Table S2 compares the measured thicknesses to the geometry optimized length of the molecules (in the gas phase, AM1 level, MOPAC). The measured thicknesses are smaller than the length (-S to terminal H). We deduce a tilt angle of ≈ 43-61° with respect to the Au surface normal for the molecules in the $^{TS}$Au-S-AZn SAMs. The area occupied per molecule is estimated from the nominal diameter of the molecules d ≈ 0.45 nm (diameter of a virtual cylinder containing the molecule plus twice the hydrogen Van der Waals radius) by $A=(\pi/4)(d/\cos\theta)^2$, which is used to estimate the number of molecules in the SAM contacted by the C-AFM tip (*vide infra*).

| | | AZ1 | AZ2 | AZ3 | AZ4 |
|---|---|---|---|---|---|
| | | 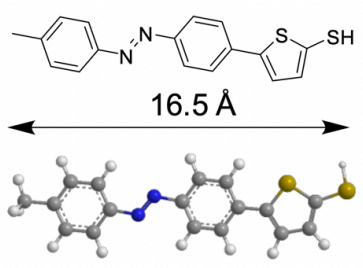  | 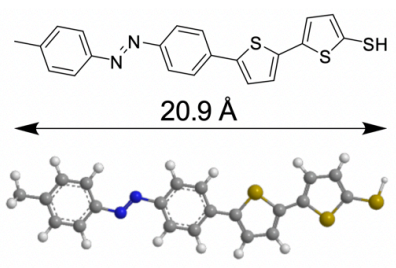  | 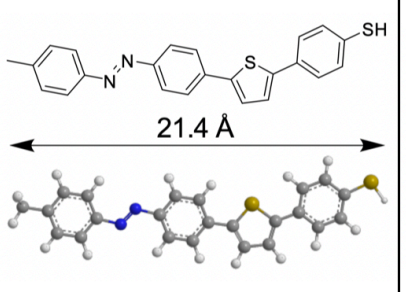  | 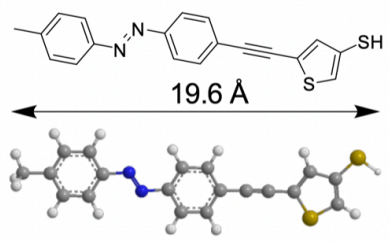  |
| Meas. thickness, t (nm) | batch #1 | 0.8 ± 0.2 | 1.0 ± 0.2 | 1.1 ± 0.2 | 1.2 ± 0.2 |
| | batch #2 | 0.96 ± 0.20 | 1.53 ± 0.20 | 1.16 ± 0.20 | 1.2 ± 0.2 |
| Theoretical length -S to H, ℓ (nm) | | 1,65 | 2,09 | 2,14 | 1,96 |
| Tilt angle, θ (°) | batch #1 | 61 | 61 | 59 | 52 |
| | batch #2 | 54 | 43 | 57 | 52 |
| Area per molecule, A ($nm^2$) | batch #1 | ≈ 0.6 | ≈ 0.6 | ≈ 0.6 | ≈ 0.4 |
| | batch #2 | ≈ 0.5 | ≈ 0.3 | ≈ 0.5 | ≈ 0.4 |

***Table S2**. Measured thickness (t) of the SAMs and comparison with the molecule length (l). The S-to-H length is taken from the geometry optimization in the gas phase shown in the upper row. We also give the tilt angle Θ from the surface normal (cosΘ = t/l) and the corresponding area per molecule occupied on the surface.*

## IV. Conductive-AFM.

### *1. General procedure.*

We measured the electron transport properties at the nanoscale by C-AFM (ICON, Bruker) at room temperature (in an air-conditioned laboratory: 22.5°C and a relative humidity of 35-40%) using a tip probe in platinum/iridium (PtIr), model SCM-PIC-V2 from Bruker. We used a "blind" mode to measure the current-voltage (*I-V*) curves and the current histograms: the C-AFM tip is randomly put in contact with the SAM and one *I-V* curve is acquired. The tip is lifted and slightly moved by ca. 50-100 nm and another *I-V* trace is acquired. This process was repeated several times at different places (randomly chosen) on the sample, and up to several thousands of *I-V* traces were used to construct the current-voltage histograms (see data sampling summary, Table S3). Notice that the C-AFM tip is regularly changed as soon as the real-time monitoring of the I-V trace indicates a significant increase in the occurrence of unreliable I-Vs as defined below (*vide infra*, "Data Analysis"). The tip load force was set at ≈ 5 nN for all the *I-V* measurements, a lower value leading to too many contact instabilities during the *I-V* measurements. Albeit larger than the usual load force (2-5 nN) used for CAFM on SAMs, this value is below the limit of about 60-70 nN at which the SAMs start to suffer from severe degradations. For example, a detailed study[10] showed a limited strain-induced deformation of the monolayer (≲ 0.3 nm) at this used load force. The same conclusion was confirmed by our own study comparing mechanical and electrical properties of alkylthiol SAMs on flat Au surfaces and tiny Au nanodots.[11]

### *2. C-AFM contact area.*

Considering: (i) the area per molecule on the surface (as estimated for the thickness measurement and calculated geometry optimization - see Table S2), and (ii) the estimated C-AFM tip contact surface (see below), we estimated the C-AFM tip contact area and the number, N, of molecules contacted by the tip as follows. As usually reported in literature[10, 12-14] the contact radius, *a*, between the C-AFM tip and the SAM surface, and the SAM elastic deformation, *δ*, are estimated from a Hertzian model:[15]

$$a^2 = \left(\frac{3RF}{4E^*}\right)^{2/3} \quad \text{(S1)}$$

$$\delta = \left(\frac{9}{16R}\right)^{1/3}\left(\frac{F}{E^*}\right)^{2/3} \quad \text{(S2)}$$

with *F* the tip load force (≈ 5 nN), *R* the tip radius (20 nm) and *E** the reduced effective Young modulus defined as:

$$E^* = \left(\frac{1}{E^*_{SAM}} + \frac{1}{E^*_{tip}}\right)^{-1} = \left(\frac{1-\nu^2_{SAM}}{E_{SAM}} + \frac{1-\nu^2_{tip}}{E_{tip}}\right)^{-1} \quad \text{(S3)}$$

In this equation, $E_{SAM/tip}$ and $\nu_{SAM/tip}$ are the Young modulus and the Poisson ratio of the SAM and C-AFM tip, respectively. For the Pt/Ir (90%/10%) tip, we have $E_{tip}$ = 204 GPa and $\nu_{tip}$ = 0.37 using a rule of mixture with the known material data.[16] These parameters for the azobenzene derivative SAMs are not known and, in general, they are not easily determined in such a monolayer material. Thus, we consider the value of an effective Young modulus of the SAM $E^*_{SAM}$ = 38 GPa as determined for the "model system" alkylthiol SAMs from a combined mechanic and electron transport study.[10] With these parameters, we estimated *a* ≈ 1.3 nm (contact area ≈ 5.5 nm$^2$) and *δ* ≈ 0.09 nm. By calculating the maximum packing of a 5.5 nm$^2$ circle by smaller circles of 0.4 nm$^2$ (on average for the $^{TS}$Au-S-AZn SAMs as estimated from the tilt angle and theoretical configuration optimization, see table S2), we infer that about 10 molecules (see www.packomania.com) are measured in the SAM/PtIr junction.

### 3. *Data analysis.*

Before constructing the current histograms and analyzing the *I-V* curves with the one energy level model, the raw set of *I-V* data was scanned and some *I-V* curves were discarded from the analysis:

- At high currents, the *I-V* traces that reached the saturating current during the voltage scan (the compliance level of the trans-impedance amplifier (here 5x10$^{-7}$ A for the used gain of the trans-impedance amplifier, marked as 'c' in Fig. S3) and/or *I-V* traces displaying large and abrupt steps (marked 'as'), Fig. S3) during the scan (contact instabilities).
- At low currents, the *I-V* traces that reached the sensitivity limit (almost flat *I-V* traces and noisy *I-V*s) and displayed random staircase behavior (due to the sensitivity limit , < 5 pA for the used gain of the trans-impedance amplifier and the resolution of the ADC (analog-digital converter), marked as noise floor 'nf' in Fig. S3.

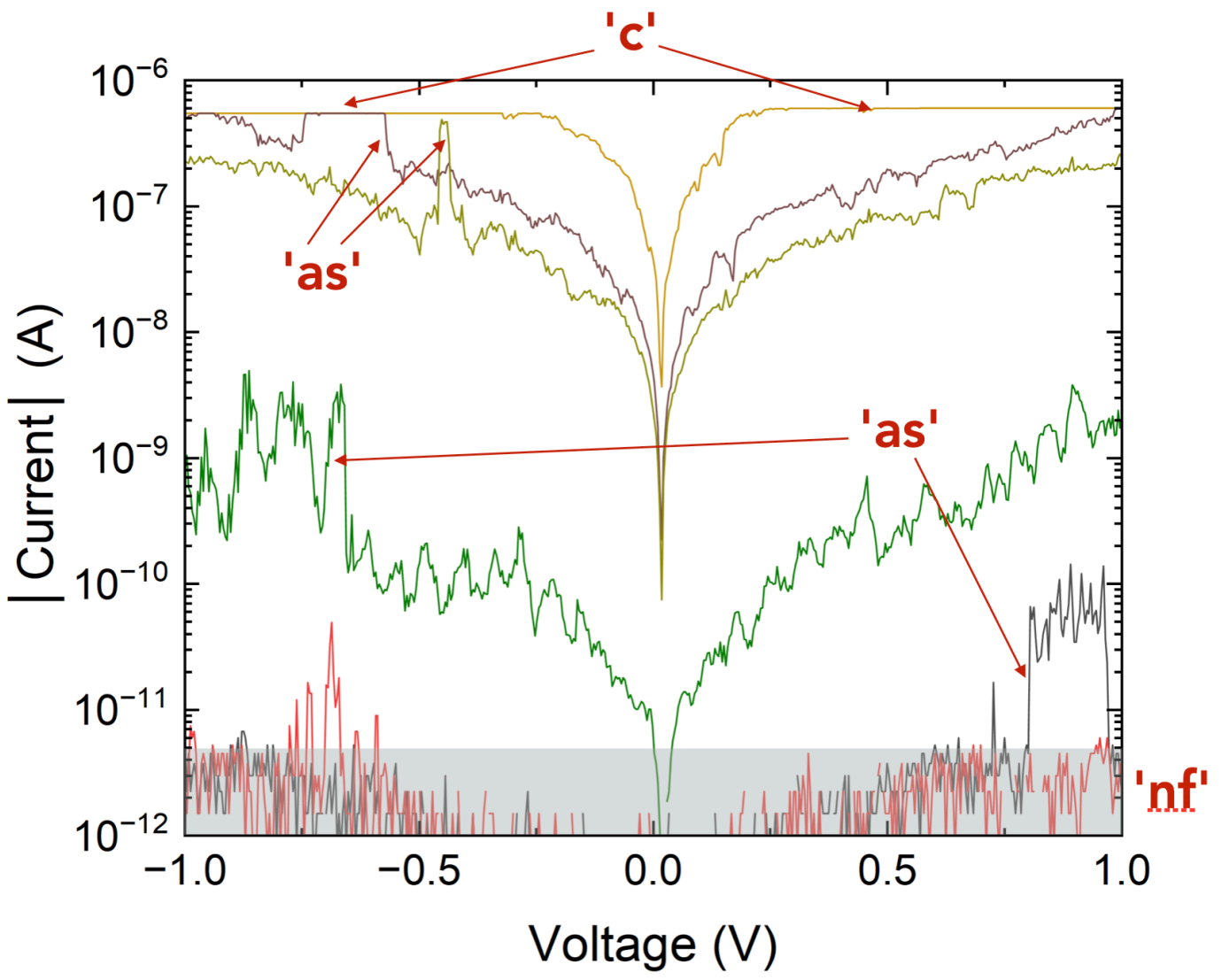

***Figure S3***. *Typical I-V traces discarded from the analysis: 'c' = compliance limit of the amplifier; 'as' = abrupt current steps (instabilities); 'nf' = noise floor limit. Data taken from batch #2, sample AZ1-pristine for the low-current data (< $10^{-9}$ A) and sample AZ1 @385 nm, 40 min for the high-level current data.*

Table S3 shows a summary of the numbers of measured I-V traces, the number of analyzed traces and the corresponding yield.

| | | batch #1 | | | | batch #2 | | | |
|---|---|---|---|---|---|---|---|---|---|
| | | # of devices | measured I-Vs | analyzed I-Vs | yield (%) | # of devices | measured I-Vs | analyzed I-Vs | yield (%) |
| AZ1 | pristine | 2 | 510 | 149 | 29,2 | 1 | 860 | 449 | 52,2 |
| | cis | | 419 | 234 | 55,8 | | 552 | 199 | 36,1 |
| | trans | | 774 | 399 | 51,6 | | 136 | 64 | 47,1 |
| AZ2 | pristine | 2 | 1078 | 289 | 26,8 | 1 | 514 | 436 | 84,8 |
| | cis | | 856 | 542 | 63,3 | | 690 | 614 | 89,0 |
| | trans | | 1106 | 289 | 26,1 | | 442 | 333 | 75,3 |
| AZ3 | pristine | 2 | 908 | 275 | 30,3 | 1 | 620 | 507 | 81,8 |
| | cis | | 765 | 265 | 34,6 | | 168 | 120 | 71,4 |
| | trans | | 1094 | 571 | 52,2 | | 314 | 292 | 93,0 |
| AZ4 | pristine | 1 | 507 | 102 | 20,1 | 2 | 600 | 348 | 58,0 |
| | cis | | 118 | 82 | 69,5 | | 158 | 116 | 73,4 |
| | trans | | 282 | 147 | 52,1 | | 118 | 64 | 54,2 |
| | | | | | 42,6 | | | | 68,0 |

***Table S3***. *Summary of the number of devices, the total number of measured I-Vs (raw data) and the number of analyzed I-Vs according to the protocol described above and the corresponding "measure yield". The last line gives the average yield for the two batches.*

## 4. Fits of the I-V curves with the analytical SEL model.

All the I-V traces of the dataset of the $^{TS}$Au-S-AZn/PtIr C-AFM tip molecular junctions (Figs. 1 and S4) were fitted with the single energy level (SEL) model given by the following analytical expression:[17, 18]

$$I(V) = N\frac{8e}{h}\frac{\Gamma_1\Gamma_2}{\Gamma_1+\Gamma_2}\left[arctan\left(\frac{\varepsilon_0+\frac{\Gamma_1}{\Gamma_1+\Gamma_2}eV}{\Gamma_1+\Gamma_2}\right)-arctan\left(\frac{\varepsilon_0-\frac{\Gamma_2}{\Gamma_1+\Gamma_2}eV}{\Gamma_1+\Gamma_2}\right)\right] \quad \text{(S4)}$$

with $\varepsilon_0$ the energy of the molecular orbital (MO), here HOMO, involved in the transport (with respect to the Fermi energy of the electrodes), $\Gamma_1$ and $\Gamma_2$ the electronic coupling energy between the MO and the electron clouds in the two electrodes, *e* the elementary electron charge, *h* the Planck constant and *N* the number of molecules contributing to the ET in the molecular junction (assuming independent molecules conducting in parallel, *i.e.*, no intermolecular interaction[19-21]). We used *N* = 10, *vide supra*. Notice that this is a crude approximation. In Eq. (S4), the values of $\Gamma_1$ and $\Gamma_2$ depend on the chosen value of N because these quantities appear in the product of the pre-factor of the arctan functions. This means that the $\Gamma_1$ and $\Gamma_2$ values are "effective" coupling energies of the SAM with the electrodes and they are used only for a relative comparison of the SAMs measured with the same C-AFM conditions in the present work and they cannot be used for a direct comparison with other reported data in the literature (as for example from single molecule experiments). On the contrary, the exact value of N has a weak influence on the fitted parameter $\varepsilon_0$, which is mainly fixed by the curvature of the I-V trace, *i.e.*, the voltage at which the I-V deviates from the linear Ohmic behavior at low bias. It is likely that, upon *trans*-to-*cis* isomerization, the number of contacted molecules can slightly change. However, even considering factor 2 in N, the variations of Γ do not change my more a factor ≈ 1.4 (Fig. S4), *i.e.*, negligible compared to the large variations (Fig. 7, main text) induced by the *trans/cis* isomerization of the azobenzene in the MJs.

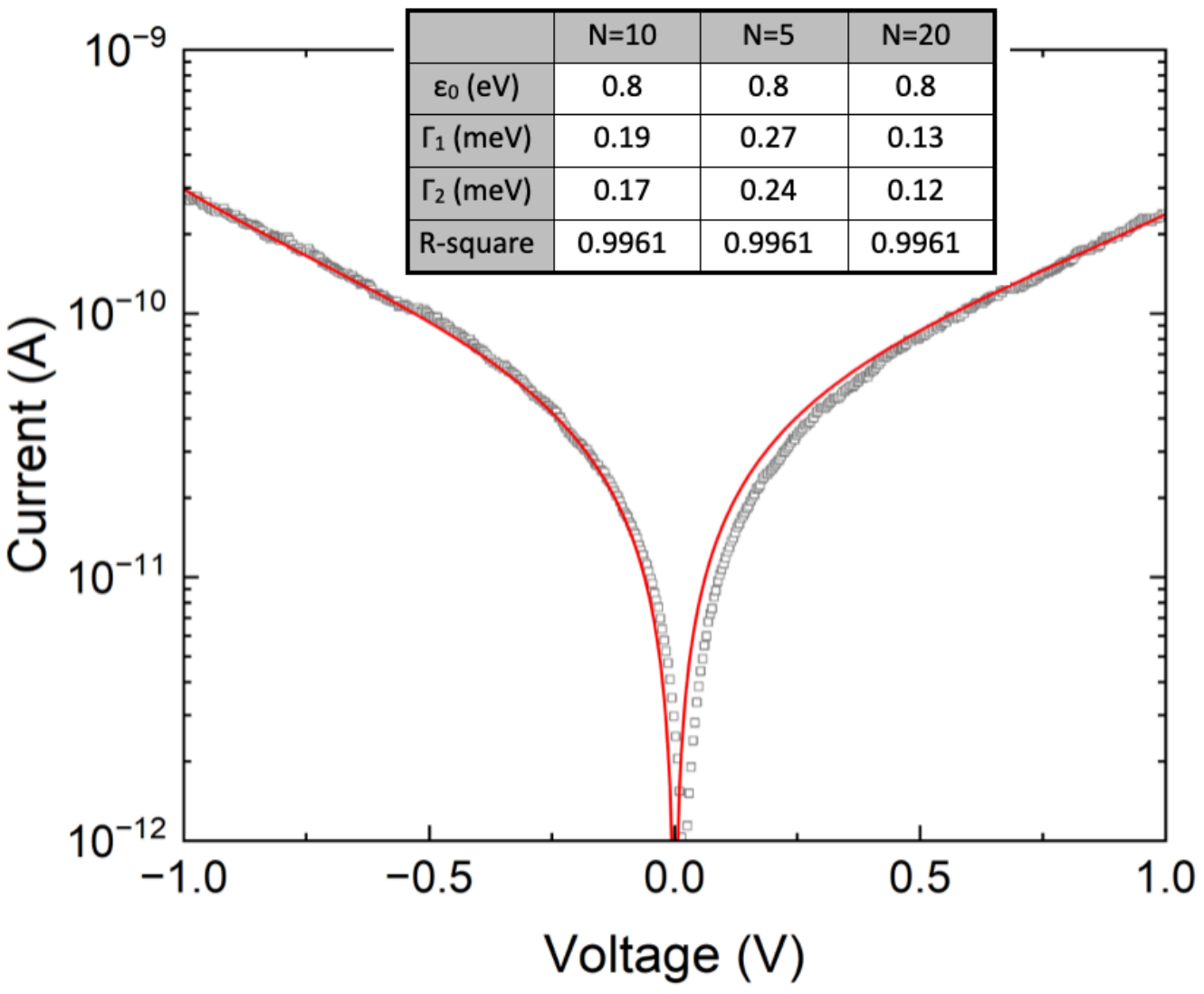

***Figure S4***. *Typical fits of the SEL model (Eq. S4) on the mean $\bar{I}$-V curve of the $^{TS}$Au-S-AZ1/tip MJ (batch 2, Fig. 1-a, main text). Compared to the 'nominal' value N=10, factor 2 induces a decrease (with N=20) or increase (with N=5) of the $\Gamma_1$ and $\Gamma_2$ values by factor ≈ 1.4. The 3 fits are superimposed.*

This model is valid at 0 K, since the Fermi-Dirac electron distribution of the electrodes is not taken into account. However, it was shown that it can be reasonably used to fit data measured at room temperature for voltages below the transition between the off-resonant and resonant transport conditions at which the broadening of the Fermi function modifies the I-V shape, leading to a sharpened increase in the current.[22-24] We defined this bias voltage window of confidence by TVS (transition voltage spectroscopy) that gives us an estimate of this transition regime. Figure S5 shows the example of the TVS curves obtained from the mean *$\bar{I}$-V* of the $^{TS}$Au-S-AZ4/tip molecular junction (batch #2). No maxima was observed in the applied voltage -1 to 1 V. Therefore, we fixed the bias window of confidence between - 1 V and 1 V to fit all the *I-Vs* of the dataset. The same conclusion holds for the three other MJs (batch #2) and the four samples of batch #1. The fits of the SEL model were done with the routine included in ORIGIN software (version 2025, OriginLab Corporation, Northampton, MA, USA), using the method of least squares and the Levenberg Marquardt iteration algorithm.

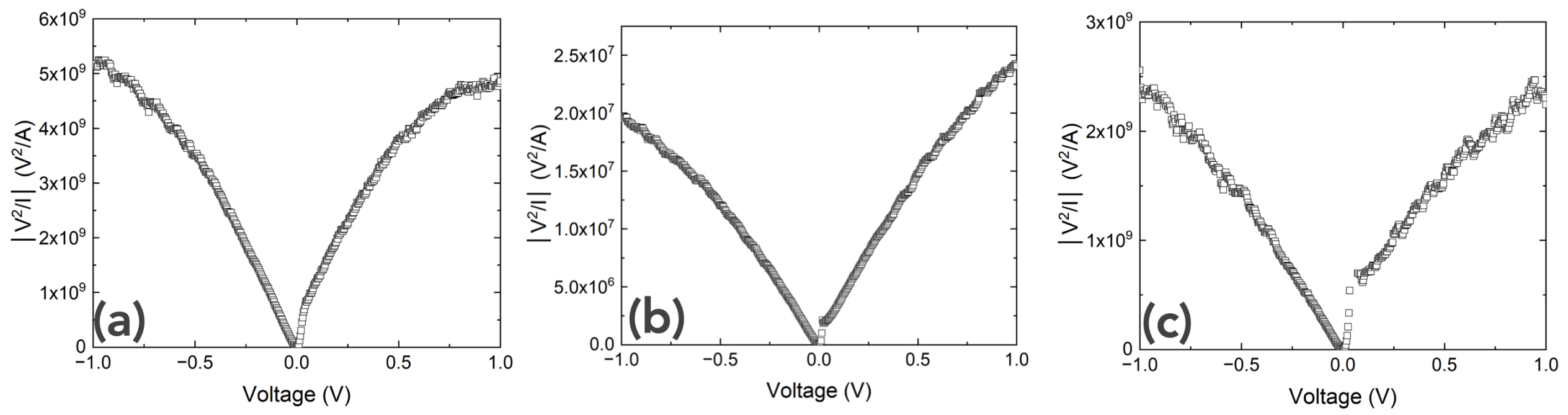

***Figure S5**. TVS plot ( |$V^2/I$| vs. V) of the mean $\bar{I}$-V of the $^{TS}$Au-S-AZ4/tip MJ molecular junction (data from Fig. 1-d,h and i, main text): **(a)** pristine, **(b)** cis-state and **(c)** trans-state.*

## V. Light irradiation.

We used LEDs (from ThorLabs) at different wavelengths coupled with a multimode optical fiber (BF20HSMA01, ThorLabs): M325F4 at 325 nm, M365FP1 at 365 nm, M385FP1 at 385 nm and M405FP1 at 405 nm, with typical FWHM bandwidths of 12 nm , 9 nm, 12 nm and 12 nm, respectively (according to the supplier data sheets : www.thorlabs.com/fiber-coupled-leds?tabName=Overview). The output of the optical fiber is placed close to the C-AFM tip (a few cm) and the light has an angle of incidence of ≈ 45°. In this geometrical configuration, the incident light intensity P is ≈ 250 µW, 1.1 mW, 1.7 mW and 2 mW at 325 nm, 365 nm, 385 nm and 405 nm, respectively (measured with a ThorLabs power meter PM204, the silicon S120VC photodetector being at the position of the sample, spot area 0.71 $cm^2$ ). All the I-V curves were measured in the dark. During light irradiation, the C-AFM tip is lifted off. The light intensity is converted into a photon flux ($\varphi$=P/h$\nu$), which is multiplied by the light irradiation time to plot the x-axis in Figs. 4-b and 4-c (main text). We also used an Analytik Jena UV lamp at 302 nm with P = 0.35 mW for some measurements on samples of batch #1.

The isomerization kinetics (Figs. 4-b and 4-c, main text) reporting the evolution of the current vs. the number of incident photons (in a log-log scale) are empirically fitted with a power law (fit parameters in Table S4) and the power-law exponents (slopes of linear regression) are used to compare the switching speed of the azobenzene in the $^{TS}$Au-S-AZn/tip MJs.

***trans* to *cis***

| | AZ1 | AZ2 | AZ3 | AZ4 |
|---|---|---|---|---|
| Intercept | -72.38 ± 7.59 | -18.61 ± 6.45 | -35.29 ± 1.24 | -64.91 ± 14.77 |
| Slope | 3.97 ± 0.41 | 1.02 ± 0.34 | 1.97 ± 0.07 | 3.56 ± 0.79 |
| Residual Sum of Squares | 0.0679 | 0.2637 | 0.004 | 0.257 |
| Pearson's r | 0.9896 | 0.8651 | 0.9983 | 0.9537 |
| R-Square (COD) | 0.9793 | 0.7485 | 0.9966 | 0.9096 |
| Adj. R-Square | 0.969 | 0.6646 | 0.9955 | 0.8644 |

***cis* to *trans***

| | AZ1 | AZ3 |
|---|---|---|
| Intercept | 75.53 ± 3.92 | 65.78 ± 13.14 |
| Slope | -4.21 ± 0.22 | -3.49 ± 0.71 |
| Residual Sum of Squares | 0.0057 | 1.178 |
| Pearson's r | -0.9986 | -0.9269 |
| R-Square (COD) | 0.9972 | 0.8591 |
| Adj. R-Square | 0.9944 | 0.8239 |

***Table S4***. *Parameters of the linear fits of the current vs. number of incident photons plotted in a log-log scale (Fig. 4, main text). The slope gives the exponent of the power law.*

## VI. Additional data, samples of batch #2.

Figure S3 shows the current ratios $R_{cis/0} = I_{cis}/I_0$ and $R_{cis/trans} = I_{cis}/I_{trans}$ versus the applied voltages calculated from the mean current-voltage ($\bar{I}$-V) shown in Figure 1 (main text).

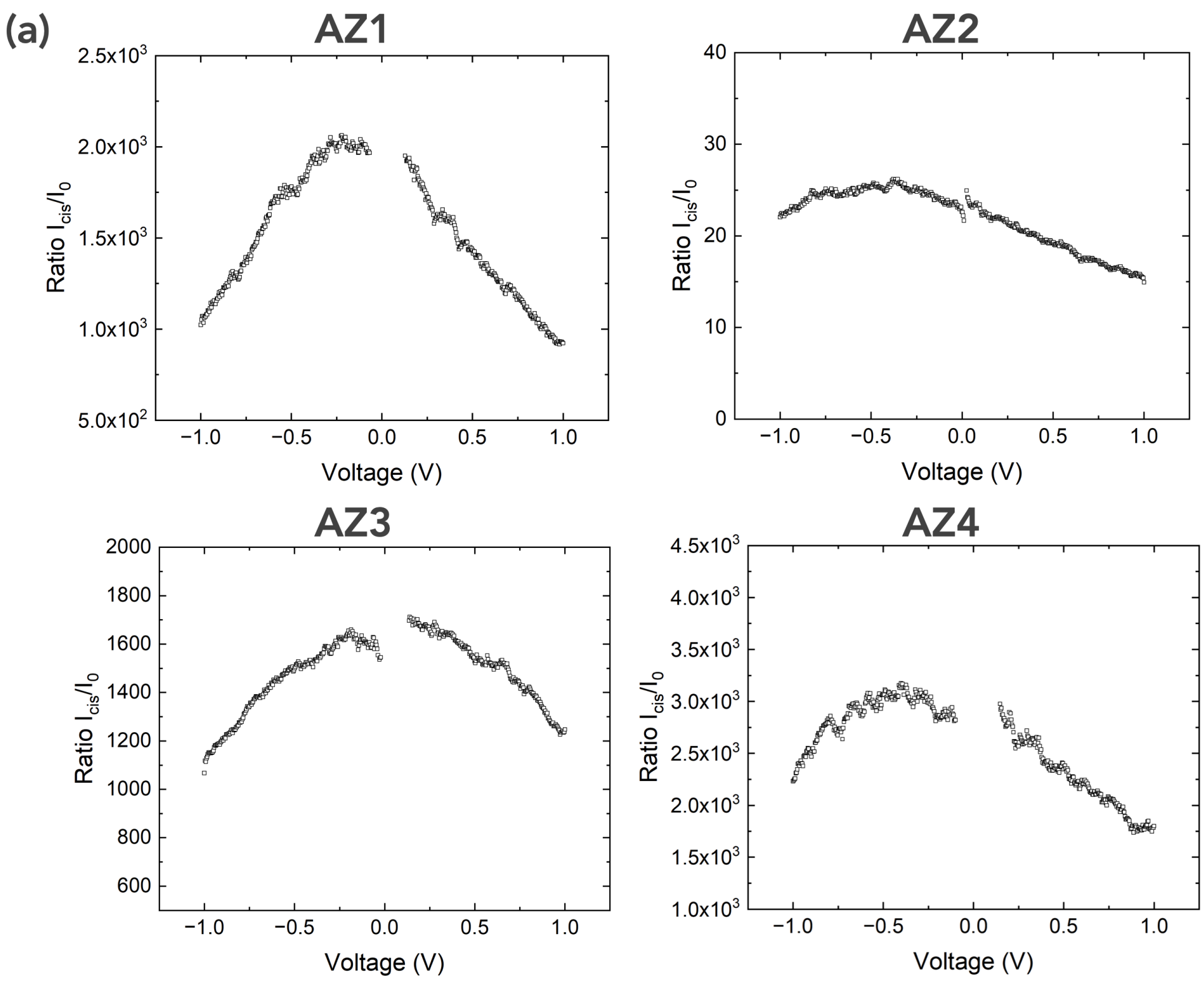

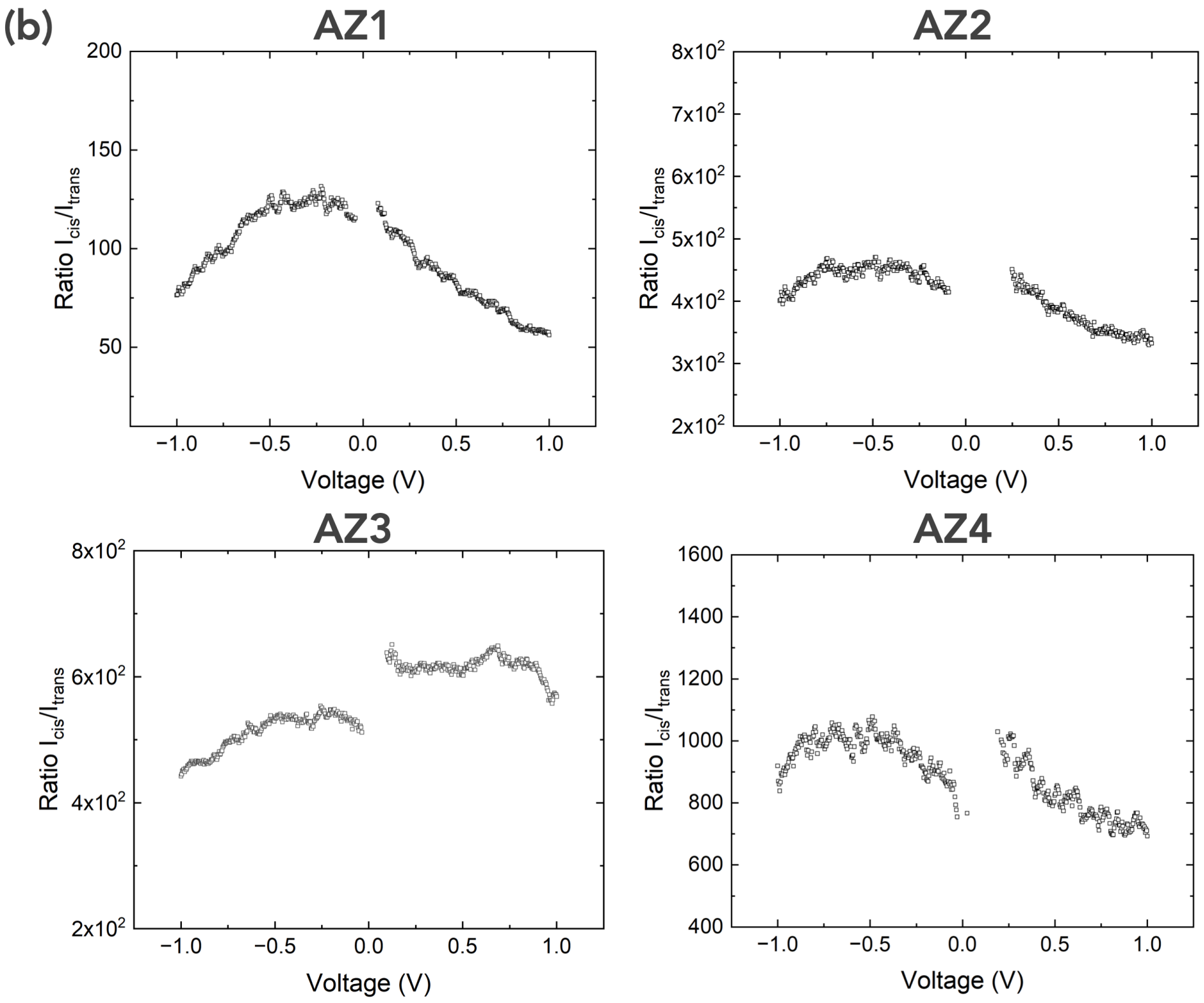

***Figure S6**. **(a)** $R_{cis/0}$ and **(b)** $R_{cis/trans}$ versus voltage calculated from the mean current-voltage ($\bar{I}$-V) shown in Figure 1 (main text).*

## VII. Data samples of batch #1.

Figure S7 shows the I-V data sets of the four $^{TS}$Au-S-AZn/tip MJs (n=1 to 4) of batch #1 fabricated and measured with the same protocol. The evolution of the SEL parameters, the HOMO energy level $\varepsilon_0$ and the electronic molecule-electrode coupling energies $\Gamma_1$ and $\Gamma_2$ are given in Figs. S9 and S10. Figure S8 gives the current histograms (at +/- 0.75 V) and the fit by a log-normal distribution (fit parameters reported in Table 1, main text).

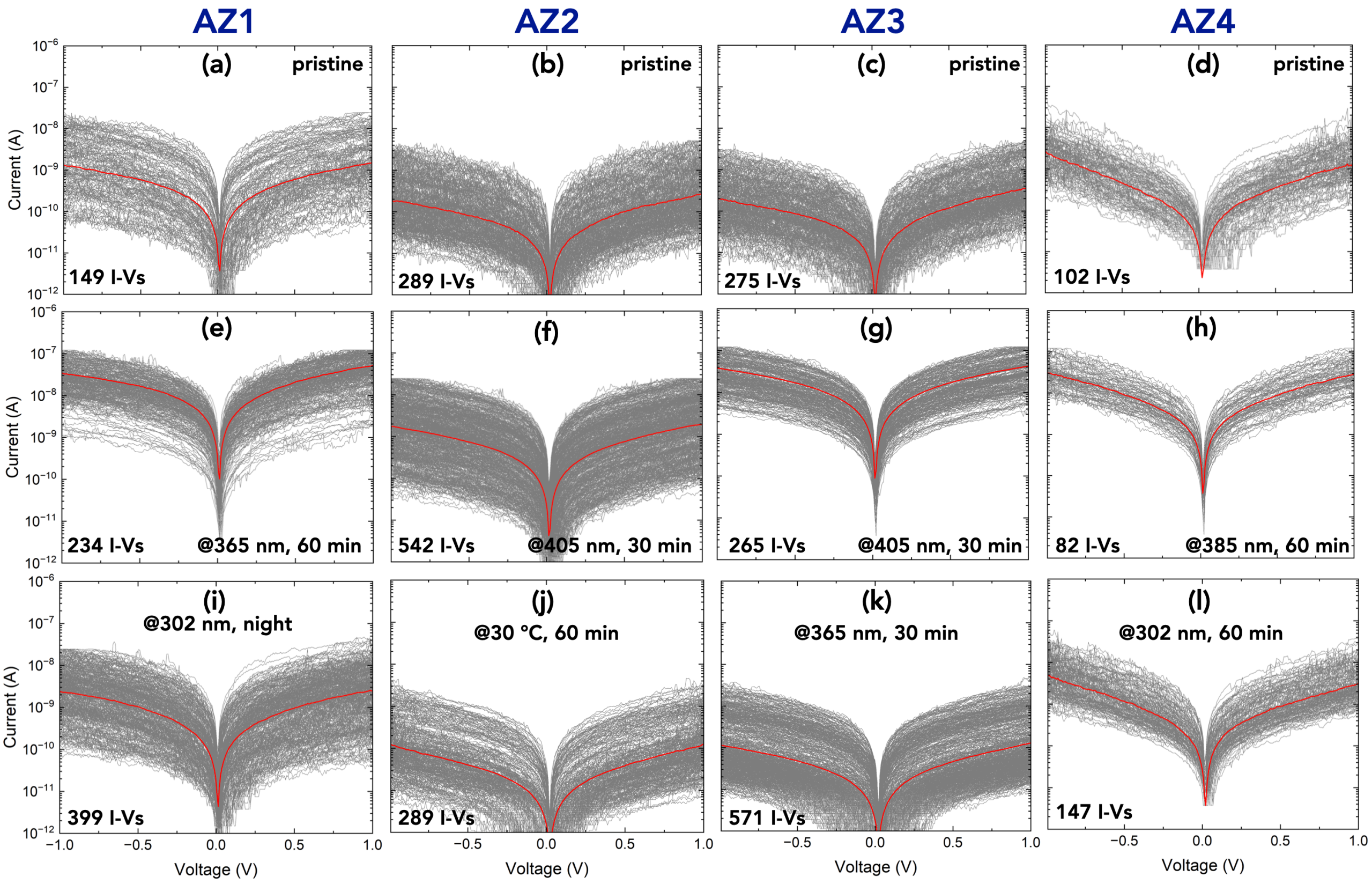

***Figure S7***. *Current-voltage (I-V) data sets for the four $^{TS}$Au-S-AZn/tip MJs (n=1 to 4, batch #1) :* ***(a-d)*** *pristine MJs after fabrication* ***(e-h)*** *after light irradiation (365-405 nm) to induce the trans-to-cis isomerization and* ***(i-l)*** *after light irradiation (302-365 nm) to induce the cis-to-trans isomerization (except for AZ2, thermal back isomerization). The red lines are the mean $\bar{I}$-V curves.*

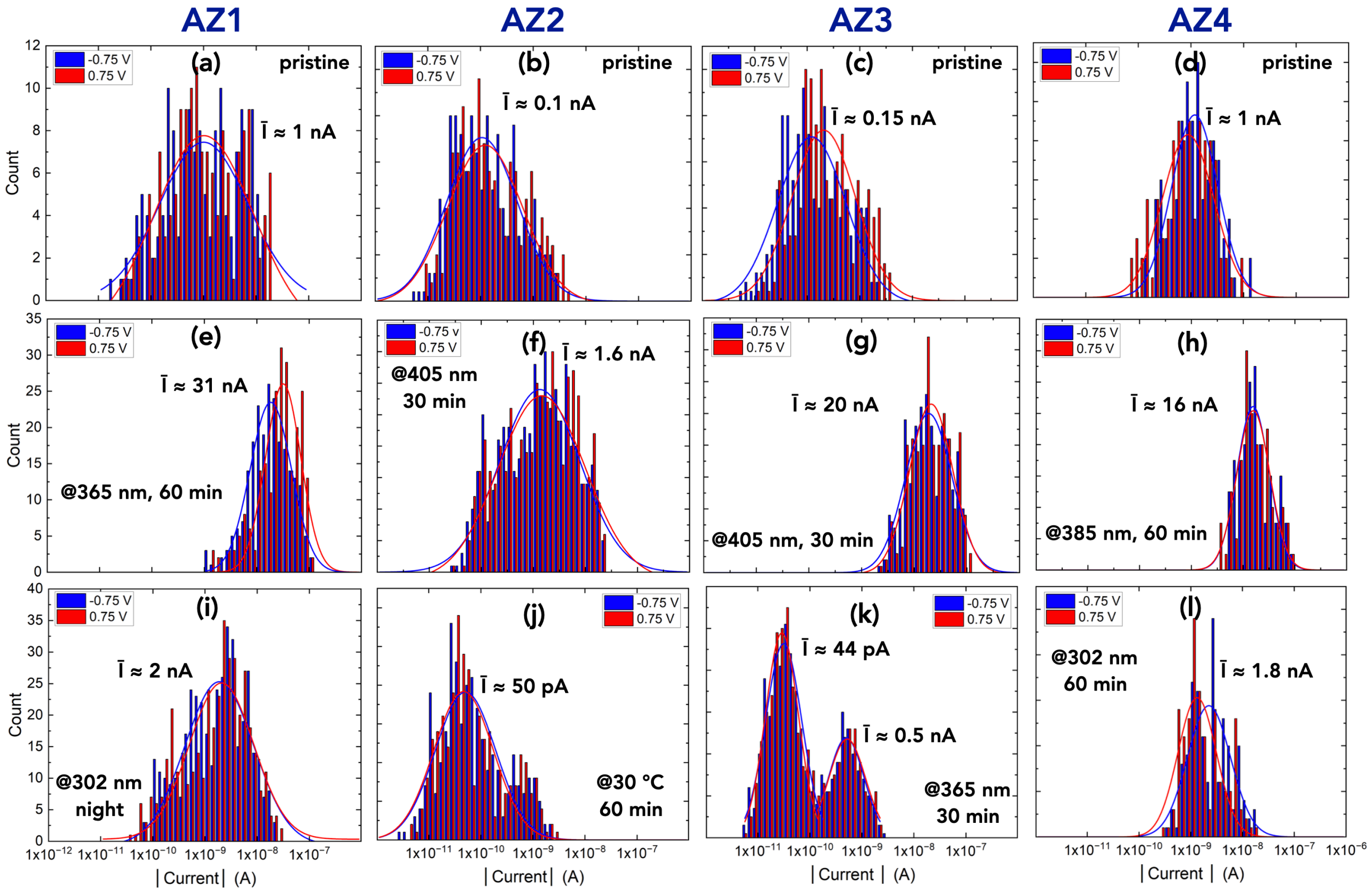

***Figure S8.*** *Histograms of the statistical distribution of the currents (at + and -0.75 V) for the four $^{TS}$Au-S-AZn/tip MJs (n=1 to 4, batch #1) :* ***(a-d)*** *pristine MJs after fabrication* ***(e-h)*** *after light irradiation (365-405 nm) to induce the trans-to-cis isomerization and* ***(i-l)*** *after light irradiation (302-365 nm) to induce the cis-to-trans isomerization (except for AZ2, thermal back isomerization). The solid lines are the fit with a log-normal distribution. The fit parameters of the log-normal distributions of the measured currents given as log-μ ± log-σ (log μ is the mean value of the Gaussian distribution in the decimal log scale, log-σ is the standard deviation in the log scale) are summarized in Table 1 (main text). The mean values Ī are given in the panels. This value is the average of the two means (at + and - 0.75 V) to cope with a small asymmetry in some cases.*

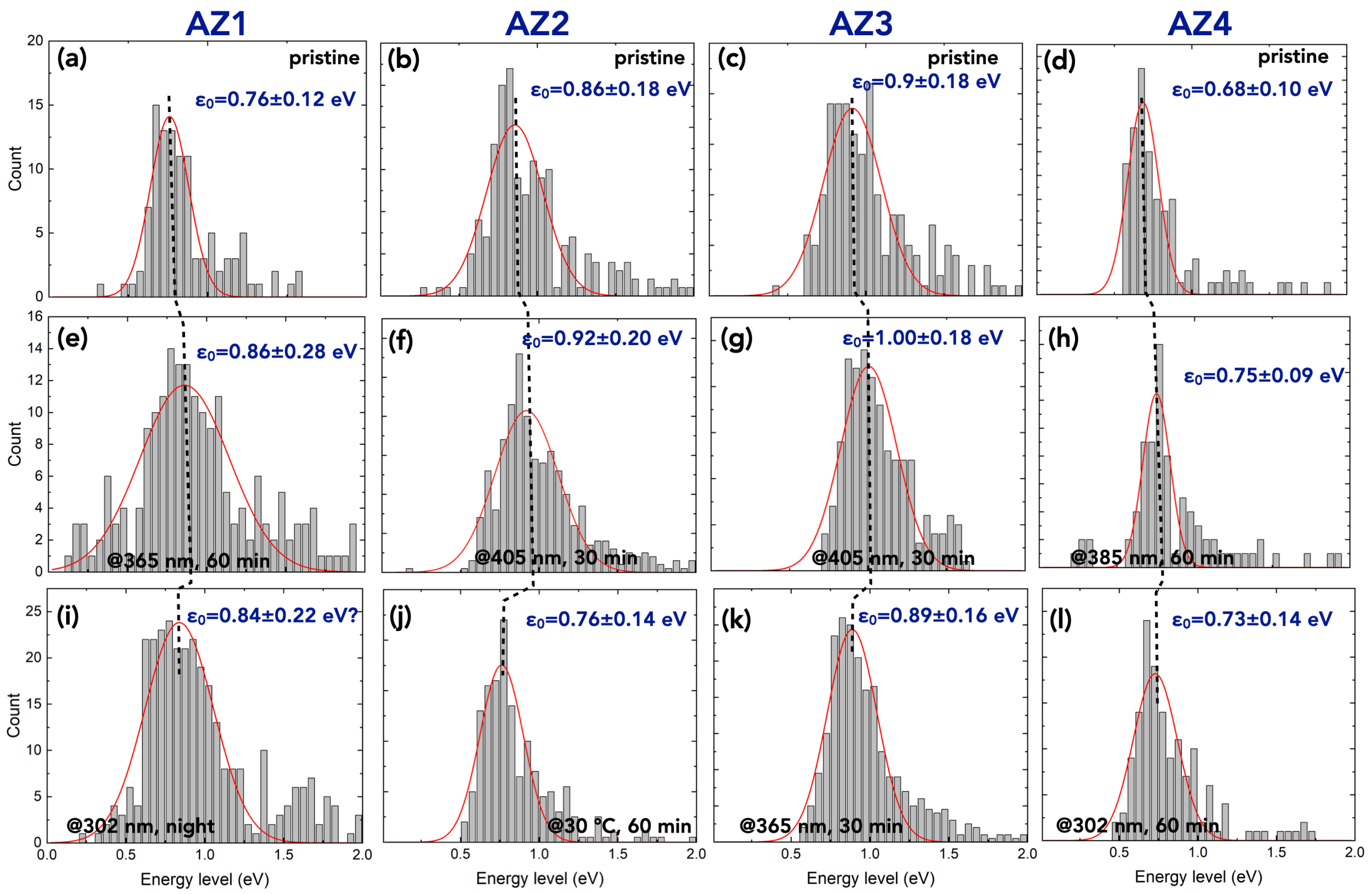

***Figure S9****. Histograms of the energy level involved in the ET though the four* $^{TS}$*Au-S-AZn/tip MJs (n=1 to 4, batch #1) :* ***(a-d)*** *pristine MJs after fabrication* ***(e-h)*** *after light irradiation (385-405 nm) to induce the trans-to-cis isomerization and* ***(i-l)*** *after light irradiation (302-365 nm) to induce the cis-to-trans isomerization (except for AZ2, thermal back isomerization). The red lines are the fits with Gaussian distributions. The mean values and the standard deviations are given in every panel.*

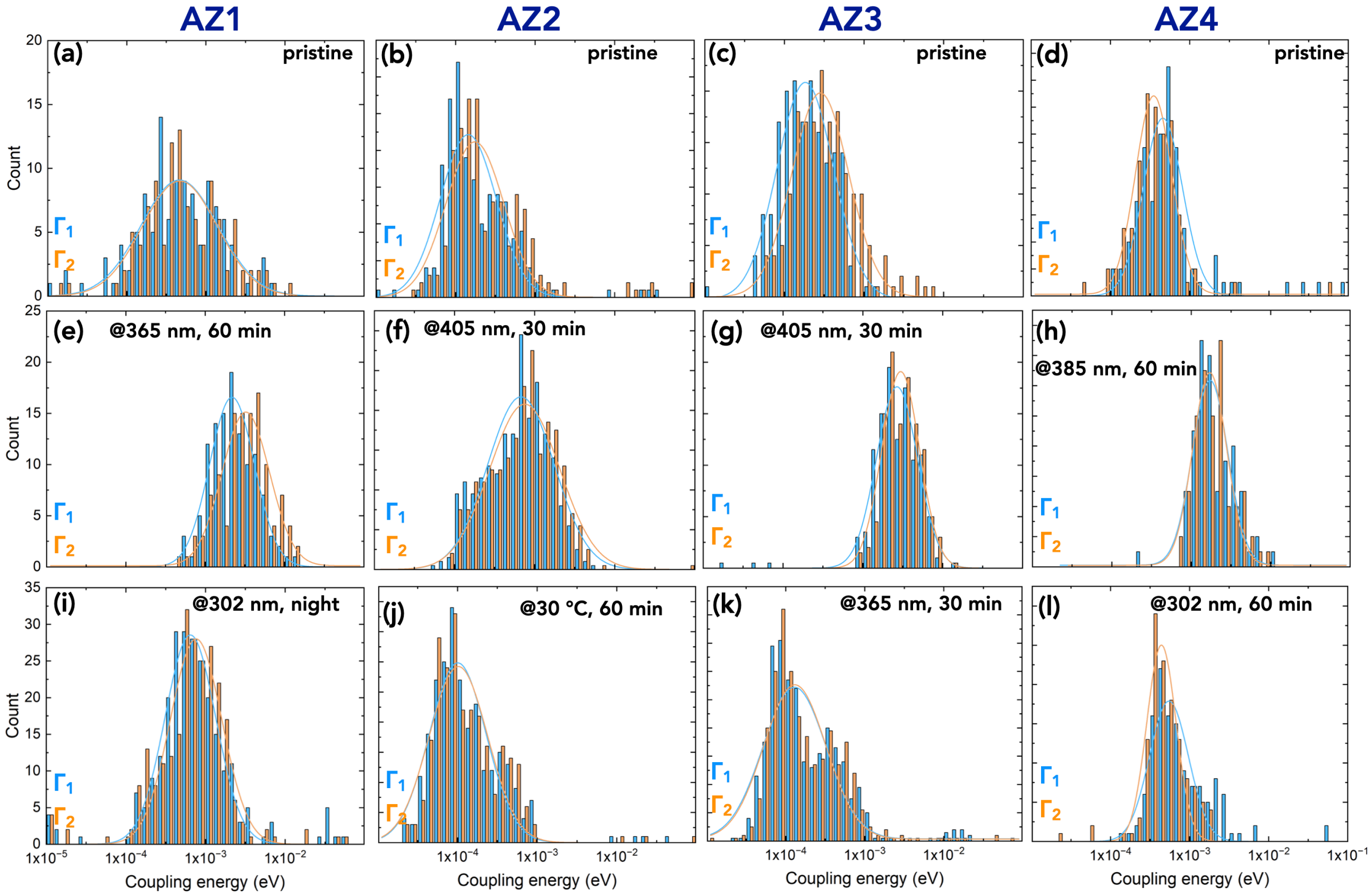

***Figure S10***. *Histograms of the molecule/electrode coupling energies ($\Gamma_1$ and $\Gamma_2$) obtained by fitting all the I-V data sets of the four $^{TS}$Au-S-AZn/tip MJs (n=1 to 4, batch #1):* ***(a-d)*** *pristine MJs after fabrication* ***(e-h)*** *after light irradiation (385-405 nm) to induce the trans-to-cis isomerization and* ***(i-l)*** *after light irradiation (302-365 nm) to induce the cis-to-trans isomerization (except for AZ2, thermal back isomerization). The solid lines are the fits with a log-normal distribution (see values Table S6). In a few cases (panels f and k) a bimodal distribution could be emerging, which has been neglected because it was not reproduced by data of batch #2 (Fig. 6 main text), and the bimodal distributions are strongly overlapping.*

## VIII. Isomerization kinetics (samples of batch #2).

Figure S11 shows the evolution of the mean Ī-V curves versus the time of light irradiation (or thermal heating for AZ2).

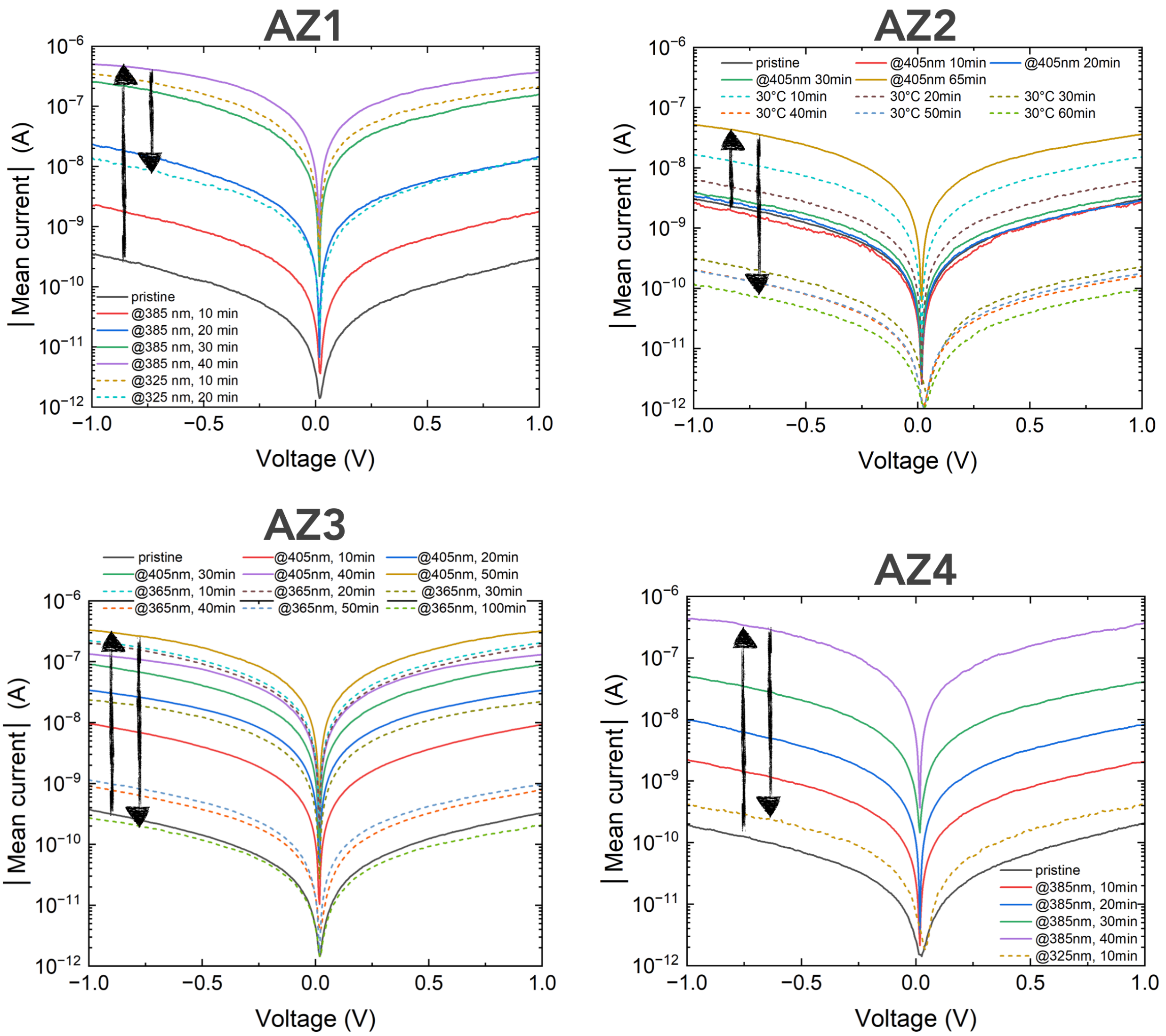

***Figure S11***. *Plot of the mean Ī-V for the four TSAu-S-AZn/tip MJs (batch #2) showing the increase in the current during the trans-to-cis isomerization (solid line I-Vs) and the decrease during the cis-to-trans isomerization (dashed line I-Vs). The wavelength and time are indicated in the panels.*

## IX. SEL model parameters.

Table S5 reports the parameters of the log-normal fits of molecule/electrode coupling energies (Figure 6, main text). Table S6 gives the same parameters for samples of batch #1 (from Fig. S10).

| | | pristine | cis | trans |
|---|---|---|---|---|
| AZ1 | $\Gamma_1$<br>$\Gamma_2$ | -3.71±0.23 (0.19)<br>-3.74±0.23 (0.18) | -2.16±0.25 (6.92)<br>-2.34±0.16 (4.57) | -2.91±0.30 (1.23)<br>-2.88±0.31 (1.32) |
| AZ2 | $\Gamma_1$<br>$\Gamma_2$ | -3.14±0.37 (0.72)<br>-3.15±0.38 (0.71) | -2.41±0.28 (3.89)<br>-2.56±0.17 (2.75) | -3.80±0.25 (0.16)<br>-3.95±0.29 (0.11) |
| AZ3 | $\Gamma_1$<br>$\Gamma_2$ | -3.56±0.32 (0.27)<br>-3.63±0.30 (0.23) | -2.20±0.18 (6.31)<br>-2.24±0.09 (5.75) | -3.34±0.43 (0.46)<br>-3.46±0.45 (0.35) |
| AZ4 | $\Gamma_1$<br>$\Gamma_2$ | -3.83±0.24 (0.15)<br>-3.83±0.22 (0.15) | -1.99±0.15 (10.23)<br>-2.15±0.07 (7.08) | -3.52±0.17 (0.30)<br>-3.55±0.17 (0.28) |

***Table S5***. *Parameters of the log-normal distributions of the coupling energies (Figure 6 main text, batch #2) given as log-μ ± log-σ (log μ is the mean value of the Gaussian distribution in the decimal log scale, log-σ is the standard deviation in the log scale). Values in brackets are the corresponding mean values of $\Gamma_1$ and $\Gamma_2$ in meV given by $10^{(log\text{-}\mu +3)}$.*

| | | pristine | cis | trans |
|---|---|---|---|---|
| AZ1 | $\Gamma_1$<br>$\Gamma_2$ | -3.33±0.48 (0.47)<br>-3.32±0.47 (0.48) | -2.49±0.30 (3.24)<br>-2.66±0.28 (2.19) | -3.19±0.32 (0.64)<br>-3.12±0.32 (0.76) |
| AZ2 | $\Gamma_1$<br>$\Gamma_2$ | -3.84±0.35 (0.14)<br>-3.76±0.36 (0.17) | -3.17±0.42 (0.68)<br>-3.12±0.44 (0.76) | -3.99±0.36 (0.10)<br>-3.98±0.37 (0.10) |
| AZ3 | $\Gamma_1$<br>$\Gamma_2$ | -3.74±0.36 (0.18)<br>-3.55±0.36 (0.28) | -2.57±0.26 (2.69)<br>-2.53±0.24 (3.39) | -3.88±0.40 (0.13)<br>-3.88±0.39 (0.13) |
| AZ4 | $\Gamma_1$<br>$\Gamma_2$ | -3.34±0.25 (0.46)<br>-3.45±0.23 (0.35) | -2.76±0.23 (1.74)<br>-2.76±0.21 (1.74) | -3.26±0.25 (0.55)<br>-3.36±0.19 (0.45) |

***Table S6***. *Parameters of the log-normal distributions of the coupling energies (Figure S10, batch #1) given as log-μ ± log-σ (log μ is the mean value of the Gaussian distribution in the decimal log scale, log-σ is the standard deviation in the log scale). Values in brackets are the corresponding mean values of $\Gamma_1$ and $\Gamma_2$ in meV given by $10^{(log\text{-}\mu +3)}$.*

## X. DFT calculations.

The geometric structure of the molecules was first optimized in the gas phase at the DFT level, with the B3LYP functional[25] and a 6-31 G (d,p) basis set[26] with the Gaussian09 software.[27] The molecules are tilted to fit the measured SAM thickness and then anchored on the gold (111) surface through a sulfur atom. The unit cell of the gold surface is modeled by a slab of six layers with 2 × 3 Au atoms in each layer and lattice parameters: a = 5.76 Å, b = 8.65 Å and α = 120°. With one molecule per unit cell, this corresponds to a theoretical density of 43.5 Å$^2$/molecule, which fits in the range of the experimental density deduced from the ellipsometry measurements (≈ 0.3-0.6 nm$^2$/molecule, see Table S2). The geometry of the interfaces was then optimized by relaxing the molecule and the top gold layers until forces are below 0.01 eV/Å. The exchange and correlation effects are described here by using the GGA.PBE exchange-correlation functional,[28] incorporating dispersion forces by Grimme correction (PBE+D2).[29] We expand the valence electrons in single zeta plus polarization (SZP) for gold atoms and a double zeta polarization basis set (DZP) for the other atoms. The core electrons are frozen and described by the norm-conserving Troullier–Martins pseudopotentials.[30] A density mesh cutoff of 100 Ha and a (4×6×1) Monkhorst Pack k-sampling were used for the optimization. Once the geometry of the interface is optimized, a layer of gold ghost atoms has been added on the top layer of the gold electrodes at a distance of 1.7 Å away so that the work function of the clean Au (111) surface of 5.25 eV matches the experimental value[31] and previous theoretical studies.[32, 33] To build a single molecular junction, we add a second gold electrode on the top side of the molecular layer by assuming a van der Waals contact, with an interatomic distance determined as the sum of Van der Waals radii of the hydrogen and gold atoms (2.86 Å). We used a gold top electrode instead of PtIr (as in the experiments) for the shake of simplicity as validated in our previous work.[34] This choice is rationalized by the fact that the work function calculated for this top electrode is 4.81 eV, in good agreement with the experimental work function of 4.86 eV of a PtIr tip.[35]

The electronic transport calculations in the junctions were performed by the combination of DFT and the Non-Equilibrium Green's Function (NEGF) formalism,[36] as implemented in QuantumATK Q-2023.12 package.[37-39] The exchange-correlation potential is described with the GGA.PBE functional[28] whereas the Brillouin zone was sampled with a (6×6×100) k-sampling. The mesh cutoff was set to 100 Ha with a temperature of 300 K. The transmission spectra at zero bias are then calculated with a 13x13 k points mesh. These parameters have been carefully tested to ensure the convergence of the transmission spectrum.

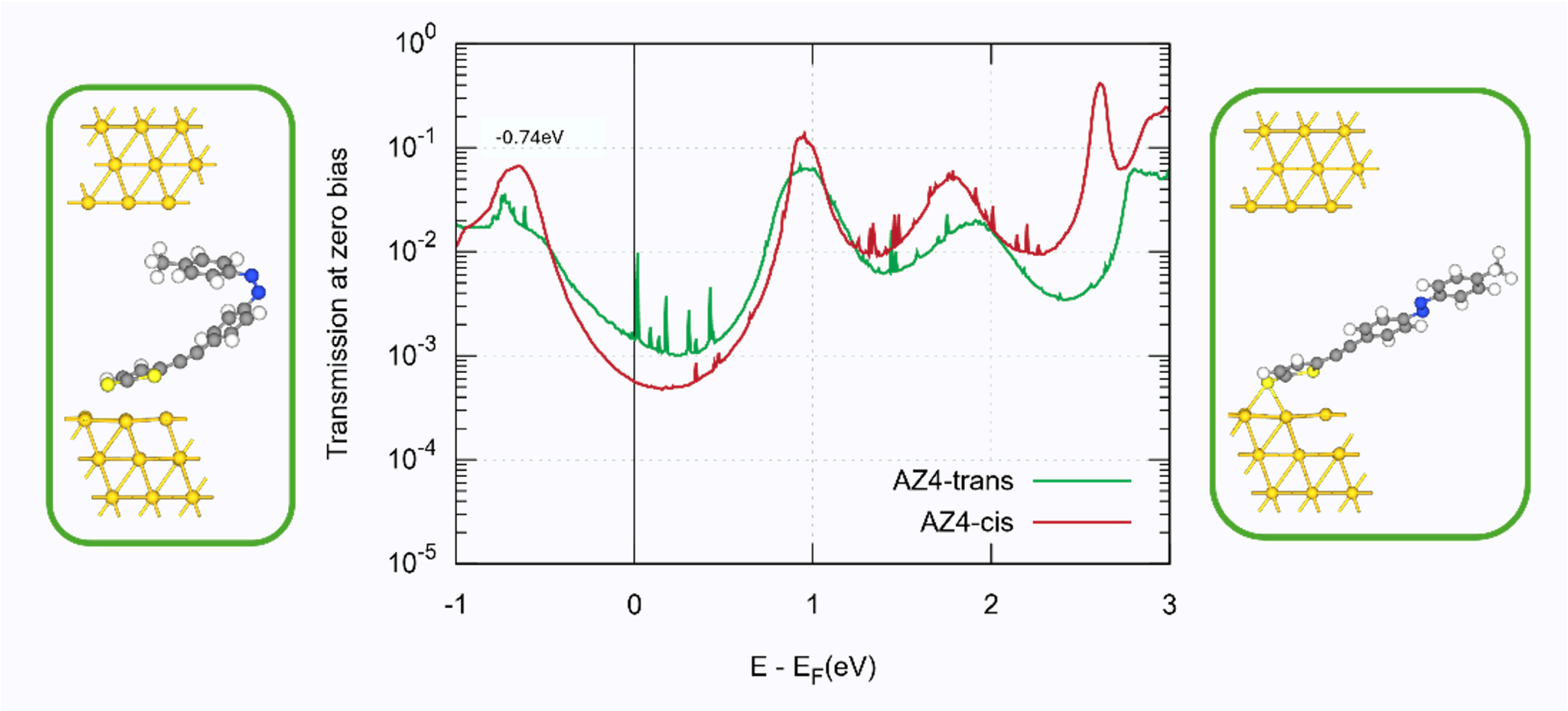

***Figure S12**. Schemes of the optimized atomic structure of the AZ4 MJ with the molecules in the trans (right panel) and cis (left panel) states and calculated electron transmission functions at zero bias. The cis/trans ratios are calculated at the Fermi energy.*

## References.

(1) Priewisch, B.; Rück-Braun, K. Efficient Preparation of Nitrosoarenes for the Synthesis of Azobenzenes. *J. Org.Chem.* **2005**, *70* (6), 2350–2352.

(2) Yamaguchi, H.; Matsuda, K. Photo- and Electrochromic Switching of Diarylethene–Gold Nanoparticle Network on Interdigitated Electrodes. *Chem. Lett.* **2009**, *38* (10), 946–947.
(3) Niebel, C.; Calard, F.; Jarrosson, T.; Lere-Porte, J. P.; Breton, T.; Serein-Spirau, F. Spontaneous assembly of silylethane-thiol derivatives on Au(111): a chemically robust thiol protecting group as the precursor for the direct formation of aromatic gold thiolate monolayers. *Chem. Commun.* **2015**, *51* (36), 7622–7625.
(4) Hnid, I.; Yassin, A.; Arbouch, I.; Guérin, D.; van Dyck, C.; Sanguinet, L.; Lenfant, S.; Cornil, J.; Blanchard, P.; Vuillaume, D. Molecular Junctions for Terahertz Switches and Detectors. *Nano Lett.* **2024**, *24* (8), 2553–2560.
(5) Hegner, M.; Wagner, P.; Semenza, G. Ultralarge atomically flat template-stripped Au surfaces for scanning probe microscopy. *Surf. Sci.* **1993**, *291*, 39–46.
(6) Weiss, E.; Chiechi, R.; Kaufman, G.; Kriebel, J.; Li, Z.; Duati, M.; Rampi, M.; Whitesides, G. Influence of defects on the electrical characteristics of mercury-drop junctions: self-assembled monolayers of n-alkanethiolates on rough and smooth silver. *J. Am. Chem. Soc* **2007**, *129* (14), 4336–4349.
(7) Weiss, E. A.; Kaufman, G. K.; Kriebel, J. K.; Li, Z.; Schalek, R.; Whitesides, G. M. Si/SiO 2-Templated Formation of Ultraflat Metal Surfaces on Glass, Polymer, and Solder Supports:  Their Use as Substrates for Self-Assembled Monolayers. *Langmuir* **2007**, *23* (19), 9686–9694.
(8) Ulman, A. *An introduction to ultrathin organic films : from Langmuir-Blodgett to Self-assembly*; Academic press, 1991.
(9) Rabe, J. P.; Knoll, W. An ellipsometric method for the characterization of macroscopically heterogeneous films. *Optics Communications* **1986**, *57* (3), 189–192.
(10) Engelkes, V. B.; Daniel Frisbie, C. Simultaneous nanoindentation and electron tunneling through alkanethiol self-assembled monolayers. *J. Phys. Chem. B* **2006**, *110* (20), 10011–10020.
(11) Smaali, K.; Desbief, S.; Foti, G.; Frederiksen, T.; Sanchez-Portal, D.; Arnau, A.; Nys, J. P.; Leclere, P.; Vuillaume, D.; Clement, N. On the mechanical and electronic properties of thiolated gold nanocrystals. *Nanoscale* **2015**, *7* (5), 1809–1819.
(12) Cui, X. D.; Primak, A.; Zarate, X.; Tomfohr, J.; Sankey, O. F.; Moore, A. L.; Moore, T. A.; Gust, D.; Harris, G.; Lindsay, S. M. Reproductible measurement of single-molecule conductivity. *Science* **2001**, *294*, 571–574.
(13) Cui, X. D.; Zarate, X.; Tomfohr, J.; Sankey, O. F.; Primak, A.; Moore, A. L.; Moore, T. A.; Gust, D.; Harris, G.; Lindsay, S. M. Making electrical contacts to molecular monolayers. *Nanotechnology* **2002**, *13*, 5–14.
(14) Morita, T.; Lindsay, S. Determination of Single Molecule Conductances of Alkanedithiols by Conducting-Atomic Force Microscopy with Large Gold Nanoparticles. *J. Am. Chem. Soc.* **2007**, *129* (23), 7262–7263.
(15) Johnson, K. L. *Contact Mechanics*; Cambridge University Press, 1987.
(16) *WebElements*. https://www.webelements.com/ (accessed 2026).
(17) Datta, S. *Electronic transport in mesoscopic systems*; Cambridge University Press, 1995.
(18) Cuevas, J. C.; Scheer, E. *Molecular Electronics: An introduction to theory and experiment*; World Scientific, 2010.
(19) Magoga, M.; Joachim, C. Conductance of molecular wires connected or bonded in parallel. *Phys. Rev. B* **1999**, *59*, 16011–16021.

(20) Reuter, M. G.; Hersam, M. C.; Seideman, T.; Ratner, M. A. Signatures of Cooperative Effects and Transport Mechanisms in Conductance Histograms. *Nano Letters* **2012**, *12* (5), 2243–2248.

(21) Trasobares, J.; Rech, J.; Jonckheere, T.; Martin, T.; Aleveque, O.; Levillain, E.; Diez-Cabanes, V.; Olivier, Y.; Cornil, J.; Nys, J. P.; et al. Estimation of π-π Electronic Couplings from Current Measurements. *Nano Letters* **2017**, *17* (5), 3215–3224.

(22) Grüter, L.; Cheng, F.; Heikkilä, T. T.; González, M. T.; Diederich, F.; Schönenberger, C.; Calame, M. Resonant tunnelling through a C 60molecular junction in a liquid environment. *Nanotechnology* **2005**, *16* (10), 2143–2148.

(23) Bourgoin, J. P.; Vuillaume, D.; Goffman, M.; Filoramo, A. Molecular Electronics. In *Nanoscience*, Dupas, C., Houdy, P., Lahmani, M. Eds.; Springer, 2007.

(24) Brunner, J. Gaining Microscopic Insight into Molecular Junctions by Transport Experiments. Basel University, 2013.

(25) Bauschlicher, C. W. A comparison of the accuracy of different functionals. *Chem. Phys. Lett.* **1995**, *246* (1), 40–44.

(26) Rassolov, V. A.; Ratner, M. A.; Pople, J. A.; Redfern, P. C.; Curtiss, L. A. 6-31G* basis set for third-row atoms. *J. Comput. Chem.* **2001**, *22* (9), 976–984.

(27) Frisch, M., et al. Gaussian 09, revision D.01. *Gaussian Inc., Wallingford CT* **2016**.

(28) Perdew, J. P.; Burke, K.; Ernzerhof, M. Generalized Gradient Approximation Made Simple. *Phys. Rev. Lett.* **1996**, *77* (18), 3865–3868.

(29) Grimme, S. Semiempirical GGA-type density functional constructed with a long-range dispersion correction. *J. Comput. Chem.* **2006**, *27* (15), 1787–1799.

(30) Troullier, N.; Martins, J. L. Efficient pseudopotentials for plane-wave calculations. *Phys. Rev. B* **1991**, *43* (3), 1993–2006.

(31) Michaelson, H. B. The work function of the elements and its periodicity. *J. Appl. Phys.* **1977**, (48), 4729–4733.

(32) Cornil, D.; Li, H.; Wood, C.; Pourtois, G.; Brédas, J.-L.; Cornil, J. Work-Function Modification of Au and Ag Surfaces upon Deposition of Self-Assembled Monolayers: Influence of the Choice of the Theoretical Approach and the Thiol Decomposition Scheme. *ChemPhysChem* **2013**, *14* (13), 2939–2946.

(33) Dyck, C.; Bergren, A. J. Large Built-In Fields Control the Electronic Properties of Nanoscale Molecular Devices with Dipolar Structures. *Adv. Electron. Mater.* **2018**, *4* (5), 1700656.

(34) Thomas, L.; Arbouch, I.; Guerin, D.; Wallart, X.; van Dyck, C.; Melin, T.; Cornil, J.; Vuillaume, D.; Lenfant, S. Conductance switching of azobenzene-based self-assembled monolayers on cobalt probed by UHV conductive-AFM. *Nanoscale* **2021**, *13* (14), 6977–6990.

(35) Lee, N. J.; Yoo, J. W.; Choi, Y. J.; Kang, C. J.; Jeon, D. Y.; Kim, D. C.; Seo, S.; Chung, H. J. The interlayer screening effect of graphene sheets investigated by Kelvin probe force microscopy. *Appl. Phys. Lett.* **2009**, *95* (22), 222107.

(36) Brandbyge, M.; Mozos, J.-L.; Ordejón, P.; Taylor, J.; Stokbro, K. Density-functional method for nonequilibrium electron transport. *Phys. Rev. B* **2002**, *65* (16), 165401.

(37) QuantumATK Q-2023.12, https://www.synopsys.com/manufacturing/quantumatk/atomistic-simulation-products.html.

(38) Smidstrup, S.; Stradi, D.; Wellendorff, J.; Khomyakov, P. A.; Vej-Hansen, U. G.; Lee, M.-E.; Ghosh, T.; Jónsson, E.; Jónsson, H.; Stokbro, K. First-principles Green's-function method for surface calculations: A pseudopotential localized basis set approach. *Phys. Rev. B* **2017**, *96* (19), 195309.
(39) Smidstrup, S.; Markussen, T.; Vancraeyveld, P.; Wellendorff, J.; Schneider, J.; Gunst, T.; Verstichel, B.; Stradi, D.; Khomyakov, P. A.; Vej-Hansen, U. G.; et al. QuantumATK: an integrated platform of electronic and atomic-scale modelling tools. *J. Phys.: Condens. Matter* **2020**, *32* (1), 015901.